\pdfoutput=1
\documentclass[11pt]{article}
\usepackage[preprint]{acl}
\usepackage{times}
\usepackage{enumitem}
\usepackage{tikz}
\usetikzlibrary{shapes, arrows.meta, positioning}
\usepackage{booktabs}
\usepackage{multirow}
\usepackage{latexsym}
\usepackage{amsmath}  
\usepackage{amssymb}   
\usepackage{tikz}
\usepackage{hyperref}
\usepackage[utf8]{inputenc}
\usepackage[T1]{fontenc}
\usepackage{amsmath,amssymb,amsthm}
\usepackage{booktabs,multirow,array}
\usepackage{graphicx}
\usepackage{stfloats}
\usepackage[font=small,labelfont=bf]{caption}
\usepackage{subcaption}
\usepackage{enumitem}
\newlength{\subfigsep}
\usepackage{algpseudocode}
\usepackage{amsmath}
\usepackage{url,float}
\usepackage{subcaption, url, float, etoolbox, adjustbox, pgf,booktabs,verbatim}
\usepackage{xcolor,graphicx,colortbl}

\definecolor{litepurple5}{RGB}{237,231,246}   % Lightest
\definecolor{litepurple10}{RGB}{209,196,233}
\definecolor{litepurple15}{RGB}{179,157,219}
\definecolor{litepurple20}{RGB}{149,117,205}
\definecolor{litepurple25}{RGB}{126,87,194}    % Darkest
\usepackage{xcolor}
\usepackage[T1]{fontenc}
\usepackage[utf8]{inputenc}
\usepackage{microtype}
\usepackage{amsmath,amssymb}      
\usepackage{algorithm}                      
\usepackage{algpseudocode}         
\usepackage{inconsolata}
\usepackage{graphicx}
\usepackage{subcaption, url, float, etoolbox, adjustbox, pgf,booktabs,verbatim}
\usepackage{xcolor,graphicx,colortbl}

\title{Towards Attribution of Generators and Emotional Manipulation in Cross-Lingual Synthetic Speech using Geometric Learning}

\author{
 \textbf{Girish\textsuperscript{1}\thanks{Equal Contribution as first authors}},
 \textbf{Mohd Mujtaba Akhtar\textsuperscript{2}\footnotemark[1]},
 \textbf{Farhan Sheth\textsuperscript{3}},
 \textbf{Muskaan Singh\textsuperscript{4}} 
\\
 \textsuperscript{1}UPES, India,
 \textsuperscript{2}V.B.S.P.U, India,
 \textsuperscript{3}Manipal University Jaipur, India,
 \textsuperscript{4}Ulster University, UK,
\\
 \small{
    \textbf{Correspondence:} \href{mailto:email@domain}
    {m.singh@ulster.uk.in}
 }
}

\begin{document}
\maketitle

\begin{abstract}
In this work, we address the problem of fine-grained traceback of emotional and manipulation characteristics from synthetically manipulated speech. We hypothesize that combining semantic–prosodic cues captured by Speech Foundation Models (SFMs) with fine-grained spectral dynamics from auditory representations can enable more precise tracing of both emotion and manipulation source. To validate this hypothesis, we introduce \texttt{MiCuNet}, a novel multitask framework for fine-grained tracing of emotional and manipulation attributes in synthetically generated speech. Our approach integrates SFM embeddings with spectrogram-based auditory features through a mixed-curvature projection mechanism that spans Hyperbolic, Euclidean, and Spherical spaces guided by a learnable temporal gating mechanism. Our proposed method adopts a multitask learning setup to simultaneously predict original emotions, manipulated emotions, and manipulation sources on the EmoFake dataset (EFD) across both English and Chinese subsets. \texttt{MiCuNet} yields consistent improvements, consistently surpassing conventional fusion strategies. To the best of our knowledge, this work presents the first study to explore a curvature-adaptive framework specifically tailored for multitask tracking in synthetic speech.
\end{abstract}

\section{Introduction and Related Work}
The rapid evolution of synthetic speech technologies has ushered in a new era of expressive and adaptable voice generation systems. Among these, emotional voice conversion (EVC) has emerged as a particularly transformative advancement, enabling the alteration of emotional tone in speech while preserving the speaker's identity and linguistic integrity.  Recent advances in synthetic speech generation have raised new challenges not only for detecting manipulated audio but also for tracing the original characteristics embedded within it. Early research efforts, notably the ASVspoof series \cite{wu15e_interspeech, kinnunen17_interspeech, todisco19_interspeech, yamagishi21_asvspoof}, laid a strong foundation for detecting spoofed speech, initially targeting synthetic speech generated via text-to-speech (TTS) and voice conversion (VC) systems, and gradually expanding to address more realistic replay attacks and deepfake scenarios. \cite{yamagishi21_asvspoof}  introduced additional challenges involving deepfake audio, compression artifacts, and mismatched telephony conditions, underscoring the growing sophistication of modern spoofing attacks. In parallel, \citet{muller2024mlaad} systematically evaluated the generalization limits of audio deepfake detectors, exposing critical vulnerabilities when models trained on controlled datasets were tested on in-the-wild samples. Beyond authenticity verification, research has increasingly shifted towards fine-grained emotion modeling. Recent works such as pre-finetuning approaches for emotional speech recognition \cite{chen23b_interspeech} and emotion prompting techniques \cite{zhou23f_interspeech} demonstrate the benefits of task adaptation and multi-task learning to capture subtle emotional cues. Emotional manipulation in synthetic speech has also gained attention, with advances in controllable emotional voice conversion \cite{qi24_interspeech} and fine-grained emotion control in voice cloning systems like EmoKnob \cite{chen2024emoknob}.   \par
% Although considerable progress has been achieved in emotion recognition and synthetic speech detection, identifying the source of manipulation in generated speech has received little attention. Recent studies have demonstrated that multilingual self-supervised models can generalize well to emotional expression across languages. 
\cite{10889008} showed that self-supervised models match human performance in cross-lingual SER, albeit with notable dialectal variability. Complementing this, \cite{upadhyay24_interspeech} proposed a layer anchoring approach to enhance generalization by aligning internal transformer layers across languages. \cite{data24} further addressed the domain mismatch using MMD-based transfer learning with 2D spectrograms, specifically targeting low-resource emotion transfer. Multitask learning (MTL) has emerged as an effective strategy to improve SER by encouraging the model to learn shared structures across related tasks. A recent MTLSER framework \cite{CHEN2025126855} co-learned emotional and auxiliary cues to improve prediction robustness, while \cite{electronics13142689} integrated speech feature enhancement into a soft decision tree–LSTM hybrid. Coordinate attention mechanisms have also been explored in multitask setups, such as in the work of \cite{SUN2025107811}, where SER accuracy was improved by adaptively focusing on salient emotional regions. The structural properties of emotion space have prompted research into non-Euclidean embeddings. 
\cite{ARANO2021115507} explore hyperbolic geometry for multimodal sentiment and emotion classification, revealing that hierarchical relationships between emotional states can be better preserved in hyperbolic space. This line of work provides a strong precedent for our own investigation into mixed-curvature manifold representations. Recent advances in emotional speech synthesis underscore the growing concern over emotionally manipulated audio. \cite{zhu2023mettsmultilingualemotionaltexttospeech} presents METTS, a multilingual text-to-speech model capable of disentangling speaker identity, language, and emotion to support high-fidelity emotional speech synthesis across speakers and languages. Extending this, 
\cite{10389638} introduces a zero-shot framework for cross-lingual emotion transfer, using predictive coding and hierarchical emotion modeling to synthesize expressive speech in unseen languages without emotional supervision. However, existing approaches largely focus either on binary detection or broad categorization, leaving the fine-grained tracebacking of original emotional states in manipulated speech largely unexplored. Addressing this critical gap, we propose \texttt{\textbf{MiCuNet}} (\textbf{Mi}xed-\textbf{Cu}rvature \textbf{Net}work) for a synthetic speech framework designed for fine-grained emotion and tracing of manipulation source from synthetic speech.  \par
Beyond authenticity verification, research has increasingly shifted towards fine-grained emotion modeling. \textit{We hypothesize that leveraging SFMs together with rich spectrogram-based auditory features has an ability. As synthetic speech systems become increasingly indistinguishable from natural speech, the need for granular interpretability and attribution in detection frameworks becomes not only relevant but imperative. For the recognition of the original emotional state, the manipulated current emotion, and the underlying manipulation source model}. To validate this hypothesis, we propose \texttt{\textbf{MiCuNet}}, a mixed-curvature feature fusion framework that systematically integrates these diverse representations under a multitask learning objective. To evaluate our hypothesis, we carry out a comprehensive comparison of state-of-the-art speech foundation models, including multilingual, monolingual, and speaker recognition pretraining paradigms. We integrate embeddings extracted from x-vector, MMS, XLS-R, Whisper, WavLM, and Wav2Vec 2.0 with spectrogram-based auditory representations obtained through STFT, CQT, and Wavelet transforms. This diverse feature integration is projected through the \texttt{\textbf{MiCuNet}} framework under a mixed-curvature multitask learning setup to enable fine-grained tracing of emotional and manipulation characteristics from synthetic speech. \newline
\noindent \textbf{To summarize, key contributions of this work are as follows:} (i) To the best of our knowledge, we are the first large-scale and systematic study exploring the effectiveness of SFMs for fine-grained emotion and manipulation source tracing from synthetically manipulated speech. Our experiments reveal that multilingual SFMs consistently outperform monolingual and speaker recognition models across both English and Chinese subsets. To this end, our work represents the first exploration of Speech foundation models for the tracebacking for the source-attributes task on the EmoFake dataset, a direction previously unexplored. (ii) We propose a novel fusion framework, \textbf{\texttt{MiCuNet}}, which synchronizes embeddings through a mixed-curvature projection into three geometric spaces such as Hyperbolic, Euclidean, and Spherical via a learnable gating mechanism. Through \textbf{\texttt{MiCuNet}}, we achieve superior performance compared to individual SFMs and standard fusion baselines, setting a new benchmark for fine-grained emotional forgery detection (EFD) in synthetic speech. \newline
\noindent \textit{All resources (code, configs, and model weights) are available at:}%
\footnote{\url{https://github.com/Helixometry/MiCuNet-IJCNLP-AACL}}

\begin{figure*}[!t]
  \centering
  %
  % Row 1: the full‐width overview
  \begin{subfigure}[b]{1.0\textwidth}
    \centering
    \includegraphics[width=\textwidth]{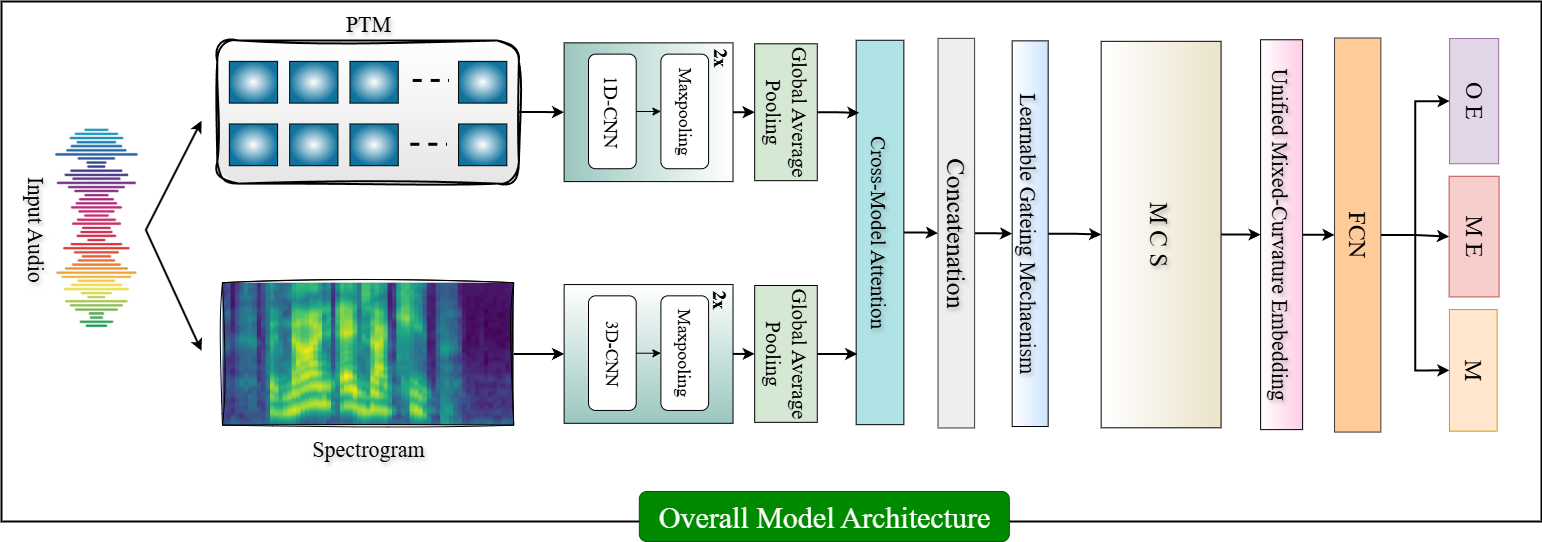}
    \caption{}
    \label{fig:1a}
  \end{subfigure}

  \vspace{1em}

  \begin{subfigure}[b]{0.29\textwidth}
    \includegraphics[width=\textwidth]{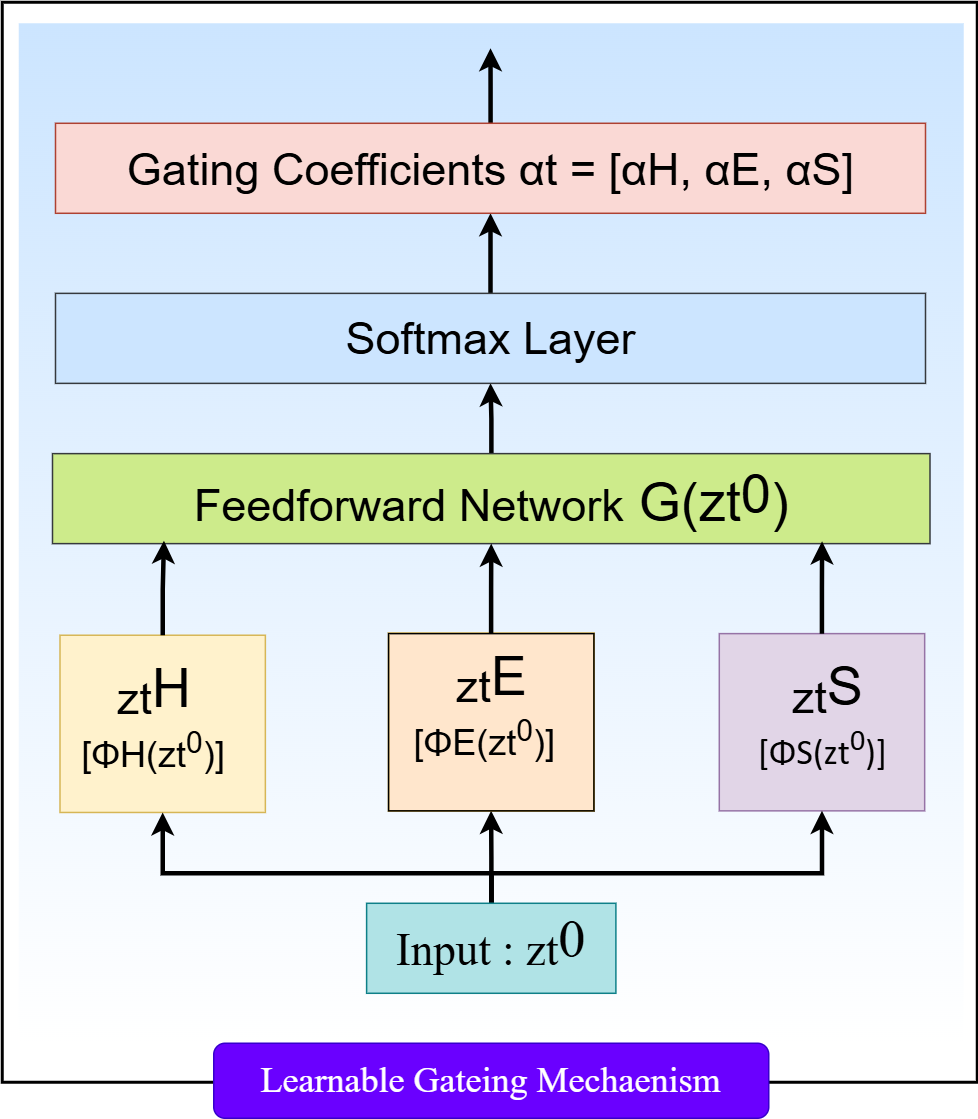}
    \caption{}
    \label{fig:1b}
  \end{subfigure}
   \hspace{\subfigsep}
  \begin{subfigure}[b]{0.28\textwidth}
    \includegraphics[width=\textwidth]{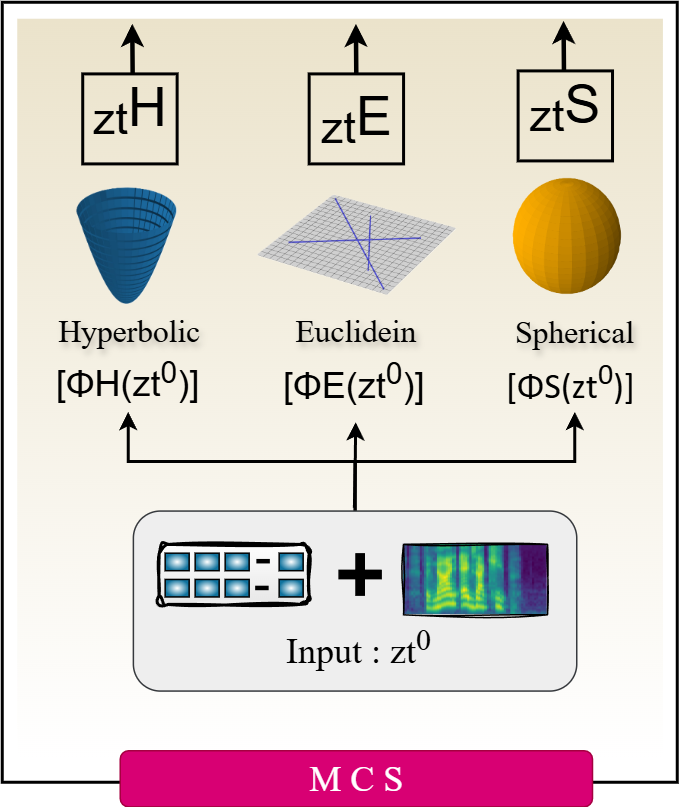}
    \caption{}
    \label{fig:1c}
  \end{subfigure}
   \hspace{\subfigsep}%
  \begin{subfigure}[b]{0.213\textwidth}
    \includegraphics[width=\textwidth]{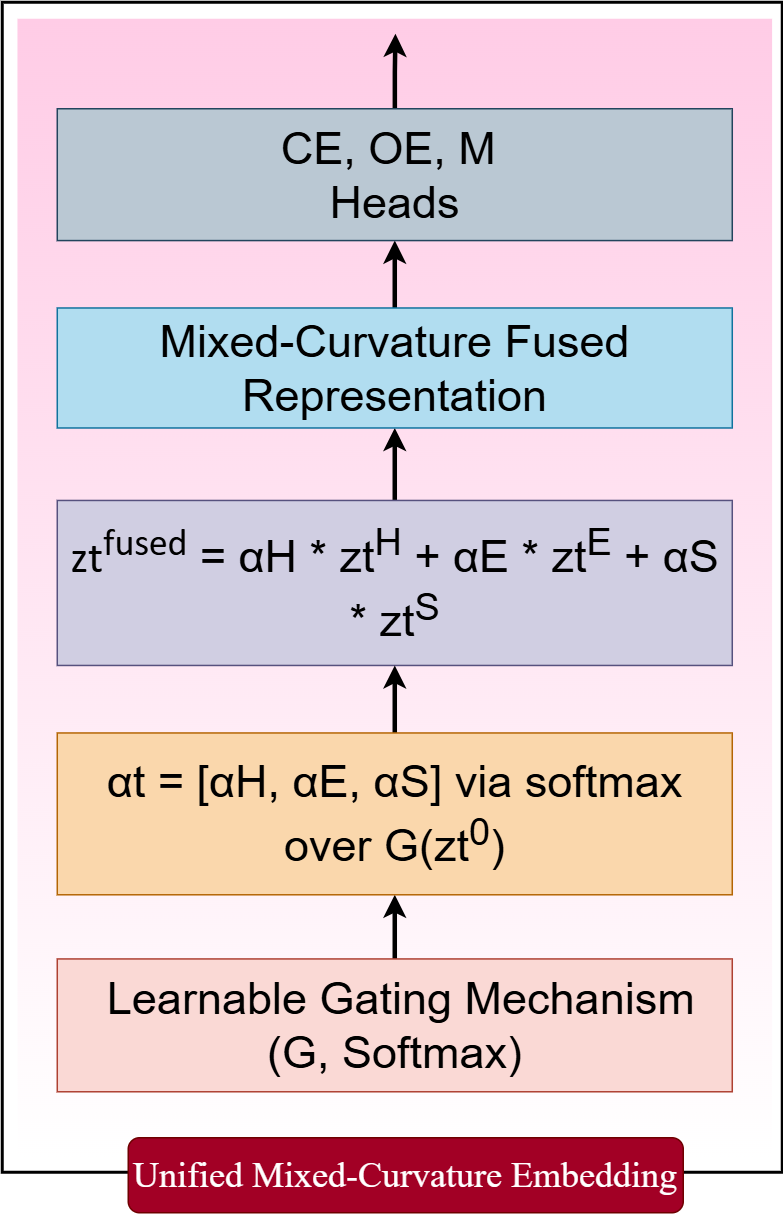}
    \caption{}
    \label{fig:1d}
  \end{subfigure}
  \caption{
    \textbf{The \textbf{\texttt{MiCuNet}} architecture.} 
    (\textbf{a}) Full pipeline: modality-specific encoders process PTM and spectrogram features using 1d and 3d CNNs, respectively, followed by self-attention, cross-modal attention, concatenation, and a curvature-aware fusion module.
    (\textbf{b}) Learnable Gating Mechanism: generates dynamic weighting coefficients for each manifold space using a feedforward network and a softmax layer.
    (\textbf{c}) Mixed-Curvature Projection: input representation is projected in parallel onto hyperbolic, Euclidean, and spherical manifolds using shared transformation layers.
    (\textbf{d}) Unified Fusion Head: weighted combination of manifold-specific embeddings followed by multitask heads for OE, CE, and M prediction.
    }
  \label{fig:1(complete-archi)}
\end{figure*}

\section{Representations}
In this section, we describe the detailed representations of the speech foundation models (SFMs) and spectrogram features adopted in our approach.

\noindent\textbf{Speech PTMs}: We use wav2vec 2.0\footnote{\url{https://huggingface.co/facebook/wav2vec2-base}} \cite{baevski2020wav2vec}, wavLM\footnote{\url{https://huggingface.co/microsoft/wavlm-base}} \cite{chen2022wavlm}, MMS\footnote{\url{https://huggingface.co/facebook/mms-1b}} \cite{pratap2024scaling}, we use XLS-R\footnote{\href{https://huggingface.co/facebook/wav2vec2-xls-r-300m}{huggingface.co/...xls-r-300m}} \cite{babu22_interspeech}, x-vector\footnote{\url{https://huggingface.co/speechbrain/spkrec-xvect-voxceleb}} \cite{8461375}, and Whisper\footnote{\url{https://huggingface.co/openai/whisper-base}} \cite{radford2023robust} as our speech foundation models. wav2vec 2.0 and wavLM are self-supervised PTMs trained on 53k and 94k hours of unlabeled audio, respectively, solving masked prediction tasks. wavLM further includes speech denoising during pretraining and improves generalization across tasks. MMS is trained on 1400 languages with 49k hours of labeled and 55k hours of unlabeled data and built on wav2vec 2.0 backbone. XLS-R extends wav2vec 2.0 to 128 languages using a multilingual contrastive learning setup. Whisper is trained in a fully supervised manner on 680k hours of web data for ASR, VAD, translation, and language ID. X-vector is a TDNN model trained on VoxCeleb1+2 for speaker verification. We use base versions of wav2vec 2.0 and wavLM (trained on English LibriSpeech). For feature extraction, we average pool the last hidden layer of all models. For Whisper, we extract from the encoder output. Audio is resampled to 16kHz. Representation dimensions are: 768 (wav2vec 2.0, wavLM), 1280 (MMS, XLS-R), 512 (x-vector, Whisper). \newline
\noindent\textbf{Spectrogram Features}: We use handcrafted features extracted from three time-frequency transforms—STFT \cite{allen1977unified}, CQT \cite{brown1991calculation}, and WT \cite{mallat1989theory}—combined with four auditory filterbanks: Mel, Gammatone, Linear, and DCT. 

\section{Modeling}
\subsection{Individual Representation Modeling}
For downstream classification, we use task-specific CNN architectures. The CNN consists of two 1d convolutional layers (64 and 128 filters, kernel size 3), each followed by ReLU activation and max pooling (pool size 2). The output is flattened and passed through a dense layer with 120 units (ReLU), followed by a softmax layer for classification. In parallel, for spectrogram-based auditory features, we implement a 3D CNN. The architecture comprises three convolutional blocks with increasing channel sizes (32 → 64 → 128). Each block includes a 3×3×3 convolution, followed by batch normalization and a ReLU activation, and is then followed by a 2×2×2 max pooling layer to reduce the spatial and temporal dimensions. \par

\begin{table*}[!hbt]
\scriptsize
\setlength{\tabcolsep}{4pt}
  \centering
  \begin{tabular}{l|ccc|ccc|ccc|ccc|ccc|ccc}
    \toprule
    \multirow{3}{*}{PTMs}
      & \multicolumn{9}{c|}{E}
      & \multicolumn{9}{c}{C} \\
    \cmidrule(lr){2-10} \cmidrule(lr){11-19}
      & \multicolumn{3}{c|}{OE}
      & \multicolumn{3}{c|}{CE}
      & \multicolumn{3}{c|}{M}
      & \multicolumn{3}{c|}{OE}
      & \multicolumn{3}{c|}{CE}
      & \multicolumn{3}{c}{M} \\
    \cmidrule(lr){2-4} \cmidrule(lr){5-7} \cmidrule(lr){8-10}
    \cmidrule(lr){11-13} \cmidrule(lr){14-16} \cmidrule(lr){17-19}
      & A & F1 & EER & A & F1 & EER & A & F1 & EER & A & F1 & EER & A & F1 & EER & A & F1 & EER \\
    \midrule
    W2   & 69.67 & 68.36 & 11.18 & 78.82 & 77.62 &  7.09 & 76.94 & 75.25 &  9.76 & 67.81 & 66.37 & 11.69 & 77.28 & 76.64 &  7.44 & 75.68 & 74.29 & 11.69 \\
    WL   & 64.05 & 62.91 & 12.88 & 75.28 & 74.61 &  9.68 & 74.52 & 74.68 & 10.09 & 63.84 & 62.51 & 13.45 & 74.67 & 73.29 & 10.17 & 72.73 & 71.68 & 13.45 \\
    MM   & 81.90 & 80.44 &  5.29 & 87.45 & 86.98 &  3.41 & 85.22 & 83.10 &  3.95 & 78.89 & 77.09 &  5.63 & 84.01 & 82.09 &  3.58 & 81.19 & 79.65 &  5.63 \\
    XL  & 79.54 & 77.12 &  6.40 & 85.32 & 84.23 &  4.09 & 81.70 & 79.16 &  4.49 & 72.49 & 71.12 &  6.76 & 82.93 & 80.99 &  4.29 & 77.69 & 76.14 &  6.76 \\
    W     & 60.93 & 59.50 & 13.56 & 65.78 & 64.40 & 10.73 & 63.20 & 62.75 &  7.29 & 59.68 & 58.24 & 14.19 & 64.81 & 63.36 & 11.28 & 62.45 & 61.89 & 14.19 \\
    XV    & 58.65 & 57.24 & 14.51 & 63.58 & 62.32 & 11.33 & 61.88 & 60.79 &  8.26 & 56.68 & 55.69 & 15.14 & 62.57 & 61.64 & 11.82 & 60.99 & 59.47 & 15.14 \\
    S-LI  & 78.56 & 77.54 &  7.41 & 81.23 & 80.12 &  3.53 & 79.82 & 78.45 &  5.47 & 77.26 & 77.72 &  7.78 & 80.03 & 79.85 &  3.71 & 78.72 & 78.20 &  7.78 \\
    S-ME & 75.23 & 73.11 &  8.05 & 76.34 & 74.32 &  5.18 & 77.41 & 75.23 &  6.29 & 74.13 & 73.28 &  8.44 & 74.94 & 74.10 &  5.40 & 76.21 & 75.05 &  8.44 \\
    S-GT  & 73.45 & 71.35 &  8.45 & 75.89 & 73.89 &  5.98 & 76.14 & 75.12 &  6.84 & 72.25 & 71.80 &  8.84 & 74.89 & 73.55 &  6.24 & 74.84 & 75.50 &  8.84 \\
    % S-DCT & 61.45 & 59.43 & 13.66 & 62.34 & 60.56 & 12.63 & 61.78 & 60.12 & 12.80 & 60.35 & 59.20 & 14.48 & 61.04 & 60.30 & 13.17 & 60.58 & 60.40 & 14.48 \\
    C-ME & 74.34 & 72.34 &  9.74 & 76.45 & 74.85 &  6.08 & 79.23 & 78.56 &  4.98 & 72.84 & 72.60 & 10.24 & 75.35 & 74.70 &  6.38 & 78.03 & 78.30 & 10.24 \\
    C-GT  & 72.89 & 70.56 &  8.77 & 77.56 & 75.47 &  5.02 & 76.78 & 75.34 &  6.31 & 71.69 & 70.10 &  9.27 & 76.26 & 75.20 &  5.24 & 75.38 & 75.60 &  9.27 \\
    C-LI  & 75.89 & 74.05 &  7.43 & 78.23 & 76.12 &  6.48 & 76.56 & 74.98 &  7.00 & 74.59 & 74.30 &  7.77 & 76.83 & 76.35 &  6.79 & 75.46 & 74.70 &  7.77 \\
    % C-DCT & 61.12 & 60.78 & 17.72 & 67.45 & 66.34 & 18.42 & 62.34 & 61.45 & 17.97 & 66.62 & 66.50 & 18.72 & 66.25 & 65.10 & 19.37 & 68.04 & 67.70 & 18.72 \\
    W-ME & 70.47 & 68.25 & 10.95 & 74.12 & 72.45 &  7.93 & 71.34 & 69.84 &  8.92 & 69.37 & 68.50 & 11.46 & 72.62 & 72.20 &  8.29 & 70.14 & 69.60 & 11.46 \\
    W-GT  & 76.34 & 74.89 &  5.85 & 79.12 & 78.23 &  5.28 & 77.56 & 75.98 &  5.66 & 75.14 & 75.10 &  6.15 & 78.12 & 78.00 &  5.55 & 76.06 & 75.75 &  6.15 \\
    W-LI  & 74.23 & 72.12 & 10.75 & 78.45 & 76.50 &  6.07 & 76.85 & 74.67 &  7.51 & 72.93 & 72.30 & 11.43 & 77.35 & 76.30 &  6.33 & 75.45 & 74.40 & 11.43 \\
    % W-DCT & 60.45 & 63.45 & 11.27 & 68.23 & 66.89 & 11.19 & 62.95 & 62.78 & 11.47 & 64.05 & 63.20 & 11.80 & 66.73 & 66.50 & 11.68 & 63.75 & 62.50 & 11.80 \\
    \bottomrule
  \end{tabular}
   \caption{Multitask performance on English (E) and Chinese (C) corpora for the following pre-trained models (PTMs); Abbreviations used: Wav2vec 2.0 (W2), WavLM (WL), MMS (MS), XLS-R (XL), Whisper (W), XVector (XV); and using spectral/time-frequency features: Constant-Q Transform–Gammatone (CQT\_GAMMATONE / C-GT), Constant-Q Transform–Linear (CQT\_LINEAR / C-LI), Constant-Q Transform–Mel (CQT\_ME / C-ME), Short-Time Fourier Transform–Gammatone (STFT\_GAMMATONE / S-GT), Short-Time Fourier Transform–Linear (STFT\_LINEAR / S-LI), Short-Time Fourier Transform–Mel (STFT\_ME / S-MEL), Wavelet Transform–Gammatone (WT\_GAMMATONE / W-GT), Wavelet Transform–Linear (WT\_LINEAR / W-LI), Wavelet Transform–Mel (WT\_ME / W-ME). Abbreviations: OE = Original–Emotion; CE = Current–Emotion; M = Model prediction; ACC = Accuracy; F1 = F1-score; EER = Equal Error Rate. \textcolor{green!70!black}{Abbreviations used are consistent across Tables:} \ref{tab:1},\ref{simpleconat},\ref{micunetconat},\ref{tab:4},\ref{tab:5},\ref{tab:6}.}
  \label{tab:1}
\end{table*}

%\vspace{-3.0mm}
\subsubsection{\texttt{MiCuNet}}
% \vspace{-4.5mm}
To better preserve nuanced emotional and manipulation cues from synthetic speech, we introduced \textbf{\texttt{MiCuNet}}, a mixed-curvature projection and fusion network designed to preserve fine-grained emotional and manipulation cues in synthetic speech. The complete system is illustrated in Figure~\ref{fig:1(complete-archi)}, where Figure ~\ref{fig:1a} depicts the full end-to-end architecture and Figures~\ref{fig:1b}–\ref{fig:1d} shows the key internal modules.

\noindent\textbf{Feature Extraction and Cross-Modal Alignment}: As illustrated in Figure~\ref{fig:1a}, we encode speech PTM embeddings using a 1D-CNN and handcrafted spectrogram features using a 3D-CNN. To obtain fixed-dimensional representations, we apply global average pooling across each modality's temporal and spatial dimensions. A cross-modal attention module enables bidirectional contextual interaction between modalities. The resulting vectors are concatenated to form the joint embedding $\mathbf{z^0}$.

\noindent\textbf{Mixed-Curvature Projection}: As shown in Figure~\ref{fig:1c}, visualizes how $\mathbf{z^0}$ is projected in parallel onto hyperbolic, Euclidean, and spherical manifolds using shared nonlinear transformations:
\begin{equation}
\mathbf{z}^\mathbb{H} = \Phi^\mathbb{H}(\mathbf{z^0}), \quad \mathbf{z}^\mathbb{E} = \Phi^\mathbb{E}(\mathbf{z^0}), \quad \mathbf{z}^\mathbb{S} = \Phi^\mathbb{S}(\mathbf{z^0})
\end{equation}
Here, $\Phi^\mathbb{H}, \Phi^\mathbb{E}, \Phi^\mathbb{S}$ are nonlinear transformation functions projecting into hyperbolic, Euclidean, and spherical spaces, respectively.

\noindent To capture the diverse structural properties inherent in emotional and manipulation signals, we leverage these three fundamental geometric manifolds:

\noindent\textbf{Hyperbolic Space ($\mathbb{H}^n$):}  
We use the Poincaré ball model of hyperbolic space with Möbius geometry to capture hierarchical and tree-like relationships prevalent in manipulated speech structures. Let $\mathbb{D}^n = \{\mathbf{x} \in \mathbb{R}^n : \|\mathbf{x}\| < 1\}$ denote the unit ball. We employ Möbius addition $\oplus$, exponential map $\exp_0^\mathbb{H}$, and logarithmic map $\log_0^\mathbb{H}$ centered at the origin. The exponential map for projecting a Euclidean tangent vector $\mathbf{v}$ onto the manifold is:

\begin{equation}
\exp_0^\mathbb{H}(\mathbf{v}) = \tanh(\sqrt{c}\|\mathbf{v}\|)\cdot \frac{\mathbf{v}}{\sqrt{c}\|\mathbf{v}\|}
\end{equation}

\noindent The corresponding logarithmic map is:

\begin{equation}
\log_0^\mathbb{H}(\mathbf{x}) = \tanh^{-1}(\sqrt{c}\|\mathbf{x}\|)\cdot \frac{\mathbf{x}}{\sqrt{c}\|\mathbf{x}\|}
\label{equa3}
\end{equation}

\noindent Here, $c$ is the curvature (learned during training), and $\|\cdot\|$ denotes the Euclidean norm. The geodesic distance used in this space is:

\begin{equation}
d_\mathbb{H}(\mathbf{x}, \mathbf{y}) = \frac{2}{\sqrt{c}} \tanh^{-1}(\|\ominus \mathbf{x} \oplus \mathbf{y}\|)
\end{equation}

\noindent This hyperbolic geometry enables \texttt{MiCuNet} to model manipulation intensities and depth transitions more effectively than Euclidean space alone.

\vspace{1mm}
\noindent\textbf{Euclidean Space ($\mathbb{E}^n$):}  
The Euclidean component of our mixed-curvature space serves as a conventional baseline for capturing linear interactions in the fused representation. No transformation is applied, and distances are measured using the $L_2$ norm:

\begin{equation}
d_\mathbb{E}(\mathbf{x}, \mathbf{y}) = \|\mathbf{x} - \mathbf{y}\|_2
\end{equation}

\noindent Euclidean embeddings complement the curvature-rich counterparts by retaining locality-sensitive structure in the feature space.

\vspace{1mm}
\noindent\textbf{Spherical Space ($\mathbb{S}^n$):}  
\noindent To capture cyclic or rotational patterns in emotional expression, we project part of the fused embedding onto a unit hypersphere of positive curvature. We use exponential and logarithmic maps centered at the north pole. The exponential map is defined as:

\begin{equation}
\exp_0^\mathbb{S}(\mathbf{v}) = \cos(\|\mathbf{v}\|)\cdot \mathbf{n} + \sin(\|\mathbf{v}\|)\cdot \frac{\mathbf{v}}{\|\mathbf{v}\|}
\end{equation}

\noindent Here, $\mathbf{n}$ denotes the north pole, and $\|\mathbf{v}\|$ is the norm of the tangent vector. The geodesic (angular) distance on the unit sphere is:

\begin{equation}
d_\mathbb{S}(\mathbf{x}, \mathbf{y}) = \arccos(\langle \mathbf{x}, \mathbf{y} \rangle)
\end{equation}

\noindent This space supports modeling speaker-intrinsic cycles and oscillatory emotional trajectories that flat or hyperbolic spaces do not capture well. \newline
\noindent By combining these spaces, the \textbf{\texttt{MiCuNet}} layer gains the capacity to model a broad spectrum of geometric dependencies within the fused representation. Instead of relying on a single manifold, MiCuNet projects the joint embedding onto three complementary geometric spaces—\textit{hyperbolic}, \textit{spherical}, and \textit{Euclidean}—each designed to capture different structural patterns in emotional speech. To fuse these manifold-specific embeddings into a unified representation, we map them into a common Euclidean space where standard arithmetic operations such as weighted summation can be applied. This transformation is achieved through logarithmic maps, which convert each curved-space embedding into its corresponding tangent space, effectively flattening the geometry for fusion. For a point \( x \) in the \textit{hyperbolic space} \( \mathbb{H}^n \), modeled via the \textit{Poincaré ball} with curvature \( c > 0 \), the logarithmic map to the Euclidean tangent space at the origin is shown in equation \ref{equa3}. \newline
% \begin{equation}
% \log_{\mathbb{H}}(x) = \tanh^{-1}(\sqrt{c} \|x\|) \cdot \frac{x}{\sqrt{c} \|x\|}
% \end{equation}
\noindent Similarly, for a point \( x \) on the \textit{unit hypersphere} \( \mathbb{S}^{n-1} \), the mapping to the tangent space at the \textit{north pole} \( n \) is:
\begin{equation}
\begin{aligned}
\log_{\mathbb{S}}(x) &= \theta \cdot \frac{x - \cos(\theta) \cdot n}{\sin(\theta)}, \\
&\text{where} \quad \theta = \arccos(\langle x, n \rangle)
\end{aligned}
\end{equation}
\noindent Once all embeddings are aligned in this shared Euclidean space, \textbf{\texttt{MiCuNet}} performs a geometry-aware fusion using a learnable gating mechanism. This module dynamically assigns weights to each manifold's contribution, allowing the model to emphasize the most informative geometry based on the task and input context. This fusion strategy not only enables richer representation learning but also enhances the model’s adaptability across different emotional, linguistic, and generative conditions. \par
\noindent\textbf{Multitask Output Heads}: As shown in Figure~\ref{fig:1d}, $\mathbf{z}_{\text{fused}}$ is passed to three independent fully-connected heads for joint prediction of Current Emotion (CE), Original Emotion (OE), and Manipulation source (M). Each head includes a hidden layer with ReLU activation followed by a softmax classifier. We specifically consider Euclidean, Hyperbolic, and Spherical spaces as they represent the three fundamental types of geometric curvature—zero, negative, and positive. Each space captures different structural relationships within the data, enabling the model to better adapt to the diverse patterns present in synthetic speech, such as linear trends, hierarchical manipulations, and cyclical emotional shifts. By jointly leveraging these complementary manifolds,  \textbf{\texttt{MiCuNet}} gains greater flexibility and robustness in representing fine-grained emotional and manipulation cues across languages. The total number of trainable parameters varies between 3.8M and 5.6M, contingent on the dimensionality of the input representations.

\section{Experiments}

\subsection{Dataset}

For our experiments, we utilize the EmoFake dataset by \citet{zhao2024emofake}\footnote{\url{https://zenodo.org/records/12806228}}. The dataset includes recordings in two languages, English and Chinese. It features five emotional states: Neutral, Happy, Angry, Sad, and Surprise. A total of 20 speakers (10 English, 10 Chinese). Synthetic (fake) emotional speech samples are created using seven publicly available EVC models: VAW-GAN-CWT, DeepEST, Seq2Seq-EVC, CycleGAN-EVC, CycleTransGAN, EmoCycleGAN, and StarGAN-EVC. Each subset contains 27,300 training samples, 9,100 development samples, and 17,500 test samples. We adopt EmoFake as it uniquely provides aligned original/manipulated speech with generator labels across two languages. The EVC systems are part of the dataset specification (not selected by us); we follow the released protocols and splits.
\newline
\noindent \textbf{Training Details}: Our models undergo 50 epochs using the Adam optimization algorithm with a starting learning rate of 0.001 and mini‑batches of 32 samples. For each of the three output tasks, we optimize the cross‑entropy objective. To mitigate overfitting, we introduce dropout, gradually reduce the learning rate over time, and stop training early if validation performance degrades. We train and evaluate the models separately on the English and Chinese subsets. All experiments are conducted on a workstation equipped with an NVIDIA RTX 3080 Ti 12 GB GPU, an Intel® Core™ i9-10900K CPU, and 64 GB of RAM.

\subsection{Experimental Results}

In this section, we present the results of our experiments, reporting performance across multiple tasks and settings.\par

\begin{table}[!hbt]
\setlength{\tabcolsep}{1.3pt}
\scriptsize
  \centering
  \begin{tabular}{l|cccccccccccc}
    \toprule
    \multirow{3}{*}{\textbf{Fusion}} 
      & \multicolumn{12}{c}{Concatenation} \\
    \cmidrule(lr){2-13}
      & \multicolumn{3}{c|}{(E-Tr) (E-Te)} 
      & \multicolumn{3}{c|}{(E-Tr) (C-Te)} 
      & \multicolumn{3}{c|}{(C-Tr) (C-Te)} 
      & \multicolumn{3}{c}{(C-Tr) (E-Te)} \\
    \cmidrule(lr){2-4}  \cmidrule(lr){5-7}  \cmidrule(lr){8-10}  \cmidrule(lr){11-13}
      & OE & CE & \multicolumn{1}{c|}{M}
      & OE & CE & \multicolumn{1}{c|}{M}
      & OE & CE & \multicolumn{1}{c|}{M}
      & OE & CE & M \\
    \midrule
    W2 + S-LI      & 3.60 & 2.15 & 3.05 & 5.16 & 3.21 & 4.32 & 3.83 & 2.26 & 3.24 & 4.62 & 2.81 & 3.98 \\
    W2 + S-ME    & 3.51 & 2.16 & 3.26 & 5.25 & 3.26 & 4.49 & 3.69 & 2.30 & 3.48 & 4.54 & 2.82 & 4.33 \\
    W2 + S-GT      & 3.52 & 2.36 & 3.39 & 5.41 & 3.38 & 4.57 & 3.74 & 2.52 & 3.58 & 4.55 & 3.04 & 4.39 \\
    W2 + C-ME     & 6.30 & 3.68 & 3.02 & 9.16 & 5.22 & 4.35 & 6.69 & 3.89 & 3.20 & 8.33 & 4.67 & 3.88 \\
    W2 + C-GT      & 4.87 & 2.82 & 3.63 & 7.05 & 4.14 & 5.11 & 5.20 & 2.98 & 3.83 & 6.33 & 3.62 & 4.77 \\
    W2 + C-LI      & 4.73 & 3.79 & 4.69 & 6.37 & 5.52 & 6.28 & 5.04 & 4.01 & 4.96 & 6.26 & 4.98 & 6.15 \\
    W2 + W-ME     & 5.48 & 4.70 & 4.81 & 7.45 & 6.36 & 7.27 & 5.79 & 4.96 & 5.09 & 7.19 & 6.13 & 6.36 \\
    W2 + W-GT      & 3.49 & 2.89 & 3.21 & 4.81 & 4.39 & 4.62 & 3.72 & 3.04 & 3.41 & 4.62 & 3.66 & 4.12 \\
    W2 + W-LI      & 5.19 & 3.13 & 4.57 & 7.27 & 4.63 & 6.27 & 5.51 & 3.29 & 4.80 & 6.80 & 4.03 & 5.98 \\
    \midrule
    WL + S-LI      & 3.82 & 2.23 & 3.16 & 5.21 & 3.25 & 4.36 & 4.02 & 2.35 & 3.32 & 4.97 & 2.93 & 4.07 \\
    WL + S-ME     & 3.93 & 2.26 & 2.98 & 5.33 & 3.32 & 4.55 & 4.15 & 2.41 & 3.18 & 5.13 & 2.94 & 3.97 \\
    WL + S-GT      & 3.62 & 2.50 & 3.43 & 5.51 & 3.44 & 4.63 & 3.83 & 2.66 & 3.64 & 4.70 & 3.26 & 4.39 \\
    WL + C-ME     & 6.87 & 3.60 & 2.97 & 9.31 & 5.31 & 4.44 & 7.35 & 3.82 & 3.16 & 8.99 & 4.67 & 3.90 \\
    WL + C-GT      & 5.08 & 2.78 & 3.66 & 7.13 & 4.19 & 5.20 & 5.36 & 2.94 & 3.91 & 6.56 & 3.63 & 4.73 \\
    WL + C-LI      & 4.41 & 3.68 & 4.43 & 6.44 & 5.61 & 6.34 & 4.71 & 3.92 & 4.68 & 5.86 & 4.76 & 5.76 \\
    WL + W-ME     & 5.13 & 4.51 & 5.43 & 7.58 & 6.43 & 7.40 & 5.44 & 4.76 & 5.76 & 6.60 & 5.84 & 7.05 \\
    WL + W-GT      & 3.19 & 3.08 & 3.18 & 4.87 & 4.45 & 4.66 & 3.40 & 3.28 & 3.40 & 4.16 & 4.02 & 4.11 \\
    WL + W-LI      & 5.52 & 3.47 & 4.49 & 7.39 & 4.71 & 6.38 & 5.87 & 3.70 & 4.78 & 7.10 & 4.55 & 5.96 \\
    \midrule
    MS + S-LI      & \cellcolor{green!25}\textbf{1.62} & \cellcolor{green!25}\textbf{1.02} & 1.45 & \cellcolor{green!25}\textbf{2.32} & \cellcolor{blue!25}\textbf{1.48} & \cellcolor{blue!25}\textbf{1.99} & \cellcolor{green!25}\textbf{1.73} & \cellcolor{green!25}\textbf{1.09} & \cellcolor{yellow!25}\textbf{1.54} & \cellcolor{green!25}\textbf{2.12} & \cellcolor{blue!25}\textbf{1.31} & 1.92 \\
    \textbf{MS + S-ME}  & \cellcolor{yellow!25}\textbf{1.64} & \cellcolor{blue!25}\textbf{1.01} & \cellcolor{green!25}\textbf{1.43} & \cellcolor{blue!25}\textbf{2.23} & \cellcolor{green!25}\textbf{1.50} & 2.10 & \cellcolor{green!25}\textbf{1.73} & \cellcolor{blue!25}\textbf{1.06} & \cellcolor{green!25}\textbf{1.52} & \cellcolor{yellow!25}\textbf{2.15} & \cellcolor{green!25}\textbf{1.32} & \cellcolor{green!25}\textbf{1.88} \\
    MS + S-GT      & 1.77 & \cellcolor{yellow!25}\textbf{1.11} & \cellcolor{blue!25}\textbf{1.38} & \cellcolor{yellow!25}\textbf{2.54} & \cellcolor{yellow!25}\textbf{1.62} & \cellcolor{yellow!25}\textbf{2.03} & \cellcolor{yellow!25}\textbf{1.87} & \cellcolor{yellow!25}\textbf{1.17} & \cellcolor{blue!25}\textbf{1.45} & 2.29 & \cellcolor{yellow!25}\textbf{1.43} & \cellcolor{blue!25}\textbf{1.78} \\
    MS + C-ME     & 3.13 & 1.55 & \cellcolor{yellow!25}\textbf{1.44} & 4.42 & 2.33 & \cellcolor{green!25}\textbf{2.02} & 3.34 & 1.64 & \cellcolor{green!25}\textbf{1.52} & 4.15 & 1.98 & \cellcolor{yellow!25}\textbf{1.89} \\
    MS + C-GT      & 2.41 & 1.37 & 1.58 & 3.34 & 1.92 & 2.35 & 2.58 & 1.46 & 1.66 & 3.11 & 1.78 & 2.02 \\
    MS + C-LI      & 2.09 & 1.73 & 1.95 & 3.00 & 2.43 & 2.85 & 2.23 & 1.84 & 2.08 & 2.72 & 2.21 & 2.58 \\
    MS + W-ME     & 2.43 & 2.06 & 2.23 & 3.36 & 2.85 & 3.29 & 2.56 & 2.20 & 2.38 & 3.16 & 2.66 & 2.97 \\
    MS + W-GT      & \cellcolor{blue!25}\textbf{1.52} & 1.45 & 1.54 & \cellcolor{blue!25}\textbf{2.23} & 2.01 & 2.18 & \cellcolor{blue!25}\textbf{1.61} & 1.54 & 1.63 & \cellcolor{blue!25}\textbf{1.99} & 1.91 & 1.96 \\
    MS + W-LI      & 2.13 & 1.47 & 1.96 & 3.27 & 2.13 & 2.80 & 2.24 & 1.56 & 2.07 & 2.72 & 1.88 & 2.50 \\
    \midrule
    XL + S-LI     & 2.34 & 1.48 & 1.99 & 3.26 & 2.07 & 2.79 & 2.48 & 1.56 & 2.11 & 2.99 & 1.92 & 2.62 \\
    XL + S-ME    & 2.37 & 1.43 & 2.01 & 3.17 & 2.02 & 2.82 & 2.52 & 1.50 & 2.12 & 3.07 & 1.83 & 2.62 \\
    XL + S-GT     & 2.33 & 1.57 & 1.90 & 3.43 & 2.19 & 2.83 & 2.48 & 1.66 & 2.02 & 3.10 & 2.01 & 2.49 \\
    XL + C-ME    & 4.05 & 2.20 & 1.97 & 5.91 & 3.29 & 2.71 & 4.27 & 2.32 & 2.10 & 5.14 & 2.81 & 2.62 \\
    XL + C-GT     & 3.11 & 1.82 & 2.32 & 4.48 & 2.56 & 3.22 & 3.31 & 1.94 & 2.44 & 4.09 & 2.35 & 2.93 \\
    XL + C-LI     & 2.90 & 2.45 & 2.94 & 4.01 & 3.43 & 3.94 & 3.05 & 2.57 & 3.12 & 3.75 & 3.20 & 3.88 \\
    XL + W-ME    & 3.13 & 2.85 & 3.26 & 4.53 & 3.83 & 4.41 & 3.34 & 2.99 & 3.43 & 4.02 & 3.72 & 4.15 \\
    XL + W-GT     & 2.13 & 1.97 & 2.06 & 3.06 & 2.77 & 2.93 & 2.28 & 2.10 & 2.17 & 2.82 & 2.52 & 2.69 \\
    XL + W-LI     & 3.32 & 2.11 & 2.57 & 4.51 & 2.87 & 3.95 & 3.55 & 2.22 & 2.73 & 4.38 & 2.73 & 3.34 \\
    \midrule
    W + S-LI        & 3.59 & 2.18 & 3.19 & 5.30 & 3.31 & 4.44 & 3.82 & 2.30 & 3.36 & 4.70 & 2.85 & 4.06 \\
    W + S-ME       & 3.71 & 2.42 & 3.32 & 5.41 & 3.35 & 4.63 & 3.93 & 2.55 & 3.55 & 4.90 & 3.18 & 4.33 \\
    W + S-GT        & 4.03 & 2.52 & 3.23 & 5.60 & 3.51 & 4.70 & 4.29 & 2.69 & 3.45 & 5.21 & 3.24 & 4.21 \\
    W + C-ME       & 7.03 & 3.75 & 3.37 & 9.46 & 5.39 & 4.52 & 7.49 & 3.97 & 3.55 & 9.00 & 4.84 & 4.31 \\
    W + C-GT        & 5.41 & 2.81 & 3.53 & 7.26 & 4.27 & 5.29 & 5.73 & 2.97 & 3.72 & 6.95 & 3.60 & 4.57 \\
    W + C-LI        & 4.59 & 4.14 & 4.53 & 6.57 & 5.70 & 6.46 & 4.90 & 4.41 & 4.82 & 5.91 & 5.47 & 5.87 \\
    W + W-ME       & 5.60 & 4.30 & 4.95 & 7.70 & 6.55 & 7.53 & 5.92 & 4.60 & 5.21 & 7.30 & 5.74 & 6.41 \\
    W + W-GT        & 3.45 & 3.19 & 3.52 & 4.95 & 4.53 & 4.75 & 3.63 & 3.35 & 3.74 & 4.47 & 4.09 & 4.64 \\
    W + W-LI        & 4.90 & 3.17 & 4.38 & 7.52 & 4.78 & 6.51 & 5.24 & 3.37 & 4.69 & 6.47 & 4.18 & 5.67 \\
    \midrule
    XV + S-LI       & 3.99 & 2.33 & 3.04 & 5.38 & 3.32 & 4.49 & 4.27 & 2.45 & 3.21 & 5.22 & 3.01 & 3.92 \\
    XV + S-ME      & 3.80 & 2.39 & 3.26 & 5.40 & 3.36 & 4.63 & 3.99 & 2.51 & 3.46 & 4.87 & 3.04 & 4.22 \\
    XV + S-GT       & 3.89 & 2.37 & 3.37 & 5.63 & 3.51 & 4.75 & 4.14 & 2.53 & 3.57 & 5.04 & 3.06 & 4.44 \\
    XV + C-ME      & 6.93 & 3.64 & 3.34 & 9.51 & 5.43 & 4.50 & 7.34 & 3.87 & 3.53 & 9.17 & 4.71 & 4.28 \\
    XV + C-GT       & 4.76 & 2.97 & 3.87 & 7.26 & 4.31 & 5.29 & 5.04 & 3.15 & 4.12 & 6.25 & 3.81 & 5.05 \\
    XV + C-LI       & 4.70 & 4.17 & 4.44 & 6.62 & 5.68 & 6.43 & 5.01 & 4.45 & 4.72 & 6.05 & 5.54 & 5.71 \\
    XV + W-ME      & 5.59 & 4.42 & 5.38 & 7.75 & 6.54 & 7.58 & 5.98 & 4.68 & 5.66 & 7.28 & 5.77 & 7.04 \\
    XV + W-GT       & 3.64 & 3.11 & 3.40 & 4.93 & 4.51 & 4.78 & 3.89 & 3.29 & 3.59 & 4.82 & 4.08 & 4.38 \\
    XV + W-LI       & 5.28 & 3.28 & 4.79 & 7.59 & 4.80 & 6.52 & 5.64 & 3.44 & 5.07 & 6.83 & 4.25 & 6.11 \\
    \bottomrule
  \end{tabular}
\caption{
Performance results using simple fusion concatenation on cross-lingual multitask prediction of Original Emotion (OE), Current Emotion (CE), and Manipulation Source (M). Results are reported as Equal Error Rates (EER \%) using 5-fold cross-validation across four train-test language combinations. Color coding in the table highlights the top three performing configurations: \textcolor{blue}{blue} (best), \textcolor{green!60!black}{green} (second best), and \textcolor{yellow!100}{yellow} (third best) based on lowest EER. consistent with the scheme used in Table~\ref{micunetconat}.}

  % \caption{Performance result using simple fusion concatenation on Cross-lingual multitask performance for CE, OE, and source (M) prediction. Results are reported as 5-fold cross in metrics Equal Error Rates (EER \%) across different train-test language combinations.}
  \label{simpleconat}
\end{table}

\noindent \textbf{Baseline Performance}: Table \ref{tab:1} presents the performance of individual pre-trained models (PTMs) and handcrafted spectral features evaluated separately on Original Emotion (OE), Current Emotion (CE), and Manipulation Source (M) across both English and Chinese corpora. Among all configurations, the Massively Multilingual Speech (MMS) model consistently delivers the strongest performance. On English CE, MMS achieves 87.45\% accuracy, 86.98\% F1, and an exceptionally low EER of 3.41\%, with similarly high scores on OE (81.90\%, 80.44\%, 5.29\%) and M (85.22\%, 83.10\%, 3.95\%). On the Chinese set, MMS maintains its lead, achieving 84.01\% accuracy and 3.58\% EER for CE, highlighting its robustness across languages and tasks. In the handcrafted domain, STFT-Linear (S-LI) emerges as a strong feature, achieving 81.23\% accuracy and 3.53\% EER on English CE, and 80.03\% and 3.71\% on Chinese CE—competitive with several PTMs. Features such as Wavelet-Gammatone (W-GT) and STFT-Mel (S-ME) also perform reliably across tasks, particularly in emotion classification, with EERs typically under 6\%. Other PTMs, including Wav2Vec 2.0, WavLM, Whisper, and X-Vector, show moderate results and provide useful contrast for understanding the impact of multilingual training, model scale, and feature richness. These patterns reinforce the importance of both cross-lingual capacity and feature diversity, motivating the need for more sophisticated fusion strategies. \newline
\noindent \textbf{Multitask Performance}: Table~\ref{simpleconat} extends the evaluation to a multitask setting using a straightforward concatenation of PTM embeddings and handcrafted features. Across all combinations, MMS paired with STFT-Mel (MS + S-ME) stands out as the top-performing setup. On English→English, it achieves 1.64\% EER (OE), 1.01\% (CE), and 1.43\% (M)—a significant improvement over single-modality baselines. In transfer settings such as English→Chinese, it maintains strong results, with EERs of 2.23\%, 1.50\%, and 2.10\%, respectively. Other combinations like MS + S-LI and MS + S-GT also perform well, consistently delivering EERs under 2.5\% across most conditions. For example, MS + S-LI reaches 1.62\%, 1.02\%, and 1.45\% on English→English, and remains competitive in cross-lingual settings. While simpler models such as Whisper or X-Vector show modest gains from concatenation, their performance remains limited compared to MMS-based pairings. These results confirm that simple fusion strategies can enhance multitask learning, especially when modalities are well-matched. However, the saturation of performance across different feature pairings indicates a need for more expressive fusion mechanisms, which \textbf{\texttt{MiCuNet}} addresses next. Table \ref{micunetconat} reports results from the proposed \texttt{MiCuNet} framework, which performs curvature-aware fusion of PTM embeddings and spectral features across three geometric manifolds. The combination of MS + S-ME achieves the best overall performance, reaching 0.66\% EER (OE), 0.31\% (CE), and 0.51\% (M) on English→English. In cross-lingual scenarios, it continues to perform exceptionally well—for instance, 1.05\%, 0.59\%, 0.71\% on English→Chinese, and 1.06\%, 0.67\, 0.95\% on Chinese→English—demonstrating robust generalization. Other fusion combinations within \texttt{MiCuNet}, such as MS + S-LI and MS + S-GT, also deliver strong performance, with EERs frequently below 1.5\% across multiple conditions. These results indicate that \texttt{MiCuNet} not only leverages the strengths of multilingual PTMs but also effectively aligns handcrafted spectral cues through its mixed-curvature projection and adaptive fusion strategy. Along with high performance the average inference time for \texttt{MiCuNet} is within range 1.2-1.6 seconds. Furthermore, the benefits of \textbf{\texttt{MiCuNet}} extend beyond the top-performing PTM. Even when applied to embeddings from models like WavLM or XLS-R, \textbf{\texttt{MiCuNet}} consistently reduces error rates, showing its flexibility across architectures and input types. \newline
\begin{table}[!ht]
\setlength{\tabcolsep}{1.3pt}
\scriptsize
  \centering
  \begin{tabular}{l|cccccccccccc}
    \toprule
    \multirow{3}{*}{\textbf{Fusion}}
      & \multicolumn{12}{c}{\texttt{MiCuNet}} \\
    \cmidrule(lr){2-13}
      & \multicolumn{3}{c|}{(E-Tr) (E-Te)}
      & \multicolumn{3}{c|}{(E-Tr) (C-Te)}
      & \multicolumn{3}{c|}{(C-Tr) (C-Te)}
      & \multicolumn{3}{c}{(C-Tr) (E-Te)} \\
    \cmidrule(lr){2-4} \cmidrule(lr){5-7} \cmidrule(lr){8-10} \cmidrule(lr){11-13}
      & OE & CE & \multicolumn{1}{c|}{M}
      & OE & CE & \multicolumn{1}{c|}{M}
      & OE & CE & \multicolumn{1}{c|}{M}
      & OE & CE & M \\
    \midrule
    W2 + S-LI   & 1.90 & 1.17 & 1.65 & 2.88 & 1.75 & 2.31 & 1.96 & 1.26 & 1.63 & 1.58 & 2.41 & 2.18 \\
    W2 + S-ME  & 1.78 & 1.17 & 1.93 & 3.14 & 1.93 & 2.32 & 1.89 & 1.20 & 1.75 & 1.47 & 2.51 & 2.26 \\
    W2 + S-GT   & 1.89 & 1.39 & 1.99 & 2.96 & 1.71 & 2.55 & 1.95 & 1.45 & 1.98 & 1.65 & 2.72 & 2.36 \\
    W2 + C-ME  & 3.55 & 2.09 & 1.64 & 4.59 & 2.90 & 2.57 & 3.91 & 2.25 & 1.71 & 2.68 & 4.61 & 2.23 \\
    W2 + C-GT   & 2.87 & 1.65 & 1.94 & 4.09 & 2.43 & 2.88 & 3.07 & 1.54 & 1.98 & 2.14 & 3.18 & 2.64 \\
    W2 + C-LI   & 2.55 & 1.90 & 2.47 & 3.64 & 3.29 & 3.69 & 2.96 & 2.29 & 2.82 & 2.78 & 3.72 & 3.26 \\
    W2 + W-ME  & 3.28 & 2.54 & 2.49 & 3.97 & 3.81 & 4.17 & 2.92 & 2.61 & 2.93 & 3.32 & 3.89 & 3.45 \\
    W2 + W-GT   & 1.76 & 1.67 & 1.78 & 2.55 & 2.26 & 2.63 & 2.08 & 1.58 & 1.75 & 1.98 & 2.36 & 2.19 \\
    W2 + W-LI   & 2.85 & 1.65 & 2.31 & 4.05 & 2.48 & 3.38 & 3.14 & 1.97 & 2.56 & 2.30 & 3.66 & 3.33 \\
    \midrule
    WL + S-LI   & 2.02 & 1.16 & 1.86 & 2.75 & 1.65 & 2.18 & 2.06 & 1.37 & 1.94 & 1.58 & 2.92 & 2.10 \\
    WL + S-ME  & 2.13 & 1.17 & 1.67 & 2.83 & 1.97 & 2.52 & 2.13 & 1.27 & 1.60 & 1.51 & 3.07 & 2.24 \\
    WL + S-GT   & 2.11 & 1.32 & 1.88 & 3.20 & 2.05 & 2.77 & 2.21 & 1.38 & 2.06 & 1.65 & 2.49 & 2.62 \\
    WL + C-ME  & 3.47 & 2.09 & 1.56 & 5.37 & 2.93 & 2.55 & 4.09 & 2.27 & 1.80 & 2.48 & 5.05 & 2.27 \\
    WL + C-GT   & 3.01 & 1.66 & 1.95 & 3.90 & 2.32 & 2.65 & 3.10 & 1.47 & 2.09 & 2.08 & 3.92 & 2.48 \\
    WL + C-LI   & 2.48 & 2.09 & 2.40 & 3.46 & 2.90 & 3.25 & 2.45 & 2.24 & 2.72 & 2.40 & 3.11 & 3.33 \\
    WL + W-ME  & 3.02 & 2.32 & 2.98 & 4.00 & 3.77 & 4.15 & 3.16 & 2.59 & 3.37 & 3.22 & 3.73 & 3.57 \\
    WL + W-GT   & 1.68 & 1.61 & 1.62 & 2.62 & 2.63 & 2.50 & 2.12 & 1.53 & 1.96 & 2.16 & 2.76 & 2.09 \\
    WL + W-LI   & 2.88 & 1.85 & 2.57 & 4.39 & 2.55 & 3.52 & 3.01 & 1.86 & 2.71 & 2.29 & 4.10 & 3.01 \\
    \midrule
    MS + S-LI   & \cellcolor{green!25}\textbf{0.77} & \cellcolor{green!25}\textbf{0.46} & \cellcolor{green!25}\textbf{0.62}
                 & \cellcolor{green!25}\textbf{1.10} & \cellcolor{green!25}\textbf{0.67} & \cellcolor{green!25}\textbf{0.86}
                 & \cellcolor{yellow!25}\textbf{0.82} & \cellcolor{yellow!25}\textbf{0.72} & \cellcolor{green!25}\textbf{0.75}
                 & \cellcolor{green!25}\textbf{0.79} & \cellcolor{green!25}\textbf{1.12} & 1.10 \\
    \textbf{MS + S-ME}
                 & \cellcolor{blue!25}\textbf{0.66} & \cellcolor{blue!25}\textbf{0.31} & \cellcolor{blue!25}\textbf{0.51}
                 & \cellcolor{blue!25}\textbf{1.05} & \cellcolor{blue!25}\textbf{0.59} & \cellcolor{blue!25}\textbf{0.71}
                 & \cellcolor{blue!25}\textbf{0.78} & \cellcolor{blue!25}\textbf{0.40} & \cellcolor{blue!25}\textbf{0.70}
                 & \cellcolor{blue!25}\textbf{0.67} & \cellcolor{blue!25}\textbf{1.06} & \cellcolor{blue!25}\textbf{0.95} \\
    MS + S-GT   & 1.05 & \cellcolor{yellow!25}\textbf{0.66} & \cellcolor{yellow!25}\textbf{0.74}
                 & 1.42 & \cellcolor{yellow!25}\textbf{0.93} & \cellcolor{yellow!25}\textbf{1.10}
                 & 0.99 & \cellcolor{green!25}\textbf{0.60} & 0.83
                 & \cellcolor{yellow!25}\textbf{0.84} & 1.36 & \cellcolor{green!25}\textbf{1.01} \\
    MS + C-ME  & 1.76 & 0.89 & 0.78 & 2.49 & 1.33 & 1.17 & 1.99 & 0.91
                 & \cellcolor{yellow!25}\textbf{0.79} & 1.17 & 2.27 & 1.08 \\
    MS + C-GT   & 1.21 & 0.74 & 0.85 & 1.86 & 1.15 & 1.29 & 1.37 & 0.80 & 0.95
                 & 1.05 & 1.74 & \cellcolor{yellow!25}\textbf{1.02} \\
    MS + C-LI   & 1.07 & 1.01 & 1.06 & 1.54 & 1.35 & 1.58 & 1.25 & 0.97 & 1.06
                 & 1.17 & 1.44 & 1.46 \\
    MS + W-ME  & 1.24 & 1.07 & 1.16 & 1.80 & 1.66 & 1.84 & 1.43 & 1.31 & 1.37
                 & 1.37 & 1.89 & 1.50 \\
    MS + W-GT   & \cellcolor{yellow!25}\textbf{0.86} & 0.82 & 0.92
                 & \cellcolor{yellow!25}\textbf{1.25} & 1.17 & 1.11
                 & \cellcolor{green!25}\textbf{0.81} & 0.91 & 0.84
                 & 1.06 & \cellcolor{yellow!25}\textbf{1.13} & 1.13 \\
    MS + W-LI   & 1.27 & 0.79 & 1.17 & 1.94 & 1.19 & 1.61 & 1.29 & 0.90 & 1.21
                 & 1.13 & 1.60 & 1.35 \\
                 \midrule
    XL + S-LI  & 1.25 & 0.85 & 1.09 & 1.78 & 1.15 & 1.59 & 1.39 & 0.83 & 1.19
                 & 0.99 & 1.52 & 1.56 \\
    XL + S-ME & 1.34 & 0.82 & 1.04 & 1.59 & 1.01 & 1.44 & 1.26 & 0.86 & 1.22
                 & 1.08 & 1.55 & 1.51 \\
    XL + S-GT  & 1.38 & 0.83 & 1.00 & 1.96 & 1.12 & 1.59 & 1.30 & 0.93 & 1.16
                 & 1.17 & 1.72 & 1.42 \\
    XL + C-ME & 2.31 & 1.30 & 1.15 & 3.47 & 1.84 & 1.46 & 2.18 & 1.30 & 1.06
                 & 1.53 & 2.94 & 1.42 \\
    XL + C-GT  & 1.74 & 0.93 & 1.26 & 2.48 & 1.47 & 1.80 & 1.82 & 1.12 & 1.39
                 & 1.32 & 2.24 & 1.54 \\
    XL + C-LI  & 1.47 & 1.39 & 1.51 & 2.30 & 2.04 & 2.19 & 1.72 & 1.52 & 1.64
                 & 1.70 & 2.02 & 2.05 \\
    XL + W-ME & 1.75 & 1.68 & 1.95 & 2.44 & 2.17 & 2.52 & 1.97 & 1.73 & 1.86
                 & 1.97 & 2.14 & 2.10 \\
    XL + W-GT  & 1.14 & 1.18 & 1.11 & 1.54 & 1.50 & 1.75 & 1.18 & 1.11 & 1.19
                 & 1.43 & 1.47 & 1.59 \\
    XL + W-LI  & 1.82 & 1.17 & 1.36 & 2.60 & 1.70 & 2.21 & 1.82 & 1.19 & 1.47
                 & 1.60 & 2.20 & 1.71 \\
                 \midrule
    W + S-LI     & 1.84 & 1.27 & 1.77 & 3.10 & 1.69 & 2.37 & 2.26 & 1.28 & 1.78
                 & 1.64 & 2.80 & 2.41 \\
    W + S-ME    & 1.89 & 1.44 & 1.75 & 2.88 & 1.71 & 2.53 & 1.97 & 1.45 & 2.00
                 & 1.76 & 2.75 & 2.17 \\
    W + S-GT     & 2.29 & 1.47 & 1.72 & 3.02 & 2.08 & 2.60 & 2.47 & 1.54 & 1.89
                 & 1.82 & 2.89 & 2.38 \\
    W + C-ME    & 3.59 & 1.88 & 1.76 & 5.36 & 2.86 & 2.65 & 4.31 & 2.36 & 1.98
                 & 2.65 & 4.70 & 2.17 \\
    W + C-GT     & 3.19 & 1.65 & 1.91 & 3.74 & 2.24 & 2.68 & 3.21 & 1.49 & 2.09
                 & 2.11 & 3.69 & 2.44 \\
    W + C-LI     & 2.71 & 2.13 & 2.49 & 3.78 & 3.09 & 3.77 & 2.91 & 2.27 & 2.67
                 & 3.22 & 3.28 & 2.95 \\
    W + W-ME    & 3.32 & 2.32 & 2.56 & 4.06 & 3.50 & 4.11 & 3.31 & 2.71 & 2.91
                 & 3.26 & 4.36 & 3.37 \\
    W + W-GT     & 1.88 & 1.88 & 1.86 & 2.50 & 2.43 & 2.80 & 2.17 & 1.72 & 2.04
                 & 2.29 & 2.45 & 2.61 \\
    W + W-LI     & 2.93 & 1.84 & 2.40 & 4.23 & 2.47 & 3.76 & 3.10 & 1.99 & 2.71
                 & 2.47 & 3.71 & 3.36 \\
                 \midrule
    XV + S-LI    & 2.02 & 1.22 & 1.82 & 3.02 & 1.69 & 2.30 & 2.14 & 1.30 & 1.68
                 & 1.70 & 2.87 & 2.20 \\
    XV + S-ME   & 2.18 & 1.35 & 1.79 & 3.09 & 1.79 & 2.54 & 2.33 & 1.48 & 2.06
                 & 1.53 & 2.92 & 2.46 \\
    XV + S-GT    & 2.30 & 1.28 & 1.72 & 2.83 & 1.85 & 2.61 & 2.26 & 1.44 & 1.81
                 & 1.68 & 2.89 & 2.27 \\
    XV + C-ME   & 3.61 & 1.83 & 1.78 & 5.50 & 3.13 & 2.63 & 4.19 & 1.96 & 1.96
                 & 2.59 & 4.79 & 2.21 \\
    XV + C-GT    & 2.65 & 1.74 & 2.00 & 4.28 & 2.23 & 3.09 & 2.99 & 1.79 & 2.30
                 & 2.24 & 3.39 & 2.91 \\
    XV + C-LI    & 2.76 & 2.25 & 2.23 & 3.71 & 2.88 & 3.70 & 2.76 & 2.56 & 2.81
                 & 2.97 & 3.38 & 3.21 \\
    XV + W-ME   & 3.13 & 2.57 & 2.83 & 4.27 & 3.58 & 4.55 & 3.55 & 2.61 & 3.28
                 & 2.92 & 4.21 & 3.62 \\
    XV + W-GT    & 2.13 & 1.62 & 1.83 & 2.60 & 2.65 & 2.56 & 2.15 & 1.71 & 1.83
                 & 2.43 & 2.75 & 2.31 \\
    XV + W-LI    & 3.08 & 1.83 & 2.82 & 4.19 & 2.49 & 3.34 & 3.15 & 1.81 & 2.85
                 & 2.33 & 3.49 & 3.22 \\
    \bottomrule
  \end{tabular}
  \caption{Performance result using proposed fusion approach \textbf{\texttt{MiCuNet}} on Cross-lingual multitask performance for CE, OE, and source (M) prediction. Results are reported as 5-fold cross in metrics Equal Error Rates (EER \%) across different train-test language combinations.}
  \label{micunetconat}
\end{table}
\begin{figure}[!t]
    \centering
    \begin{subfigure}{0.3\columnwidth}
        \includegraphics[width=\linewidth]{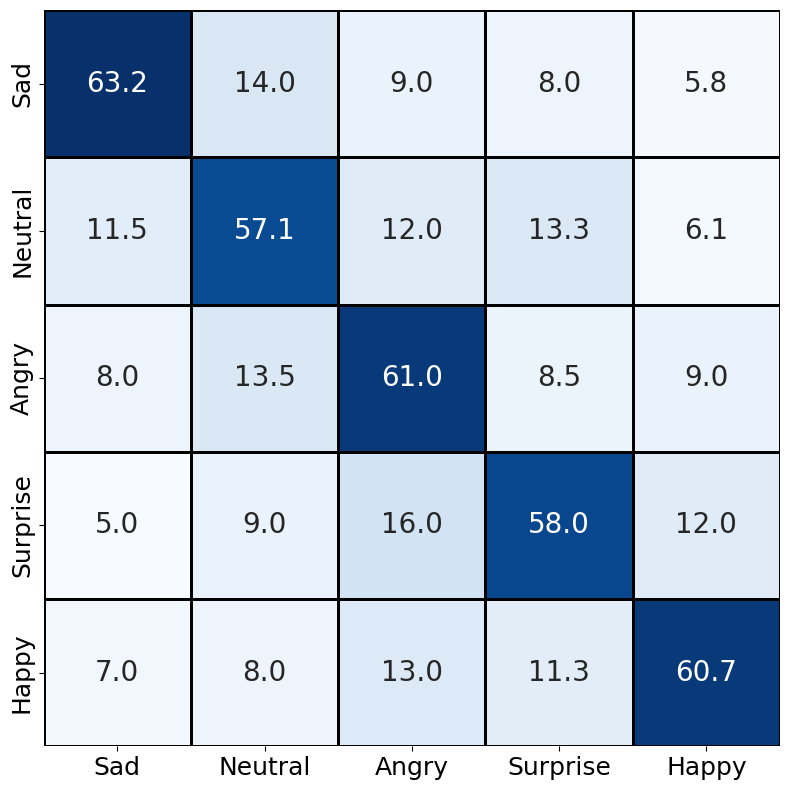}
        \caption{}
        \label{cma}
    \end{subfigure}
    \hfill
    \begin{subfigure}{0.3\columnwidth}
        \includegraphics[width=\linewidth]{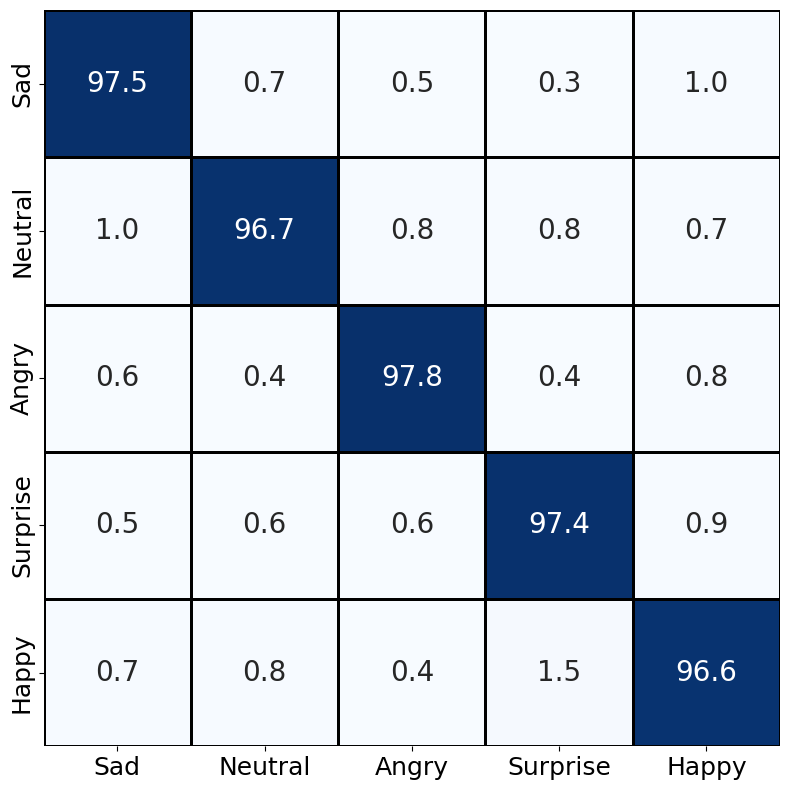}
        \caption{}
        \label{cmb}
    \end{subfigure}
    \hfill
    \begin{subfigure}{0.3\columnwidth}
        \includegraphics[width=\linewidth]{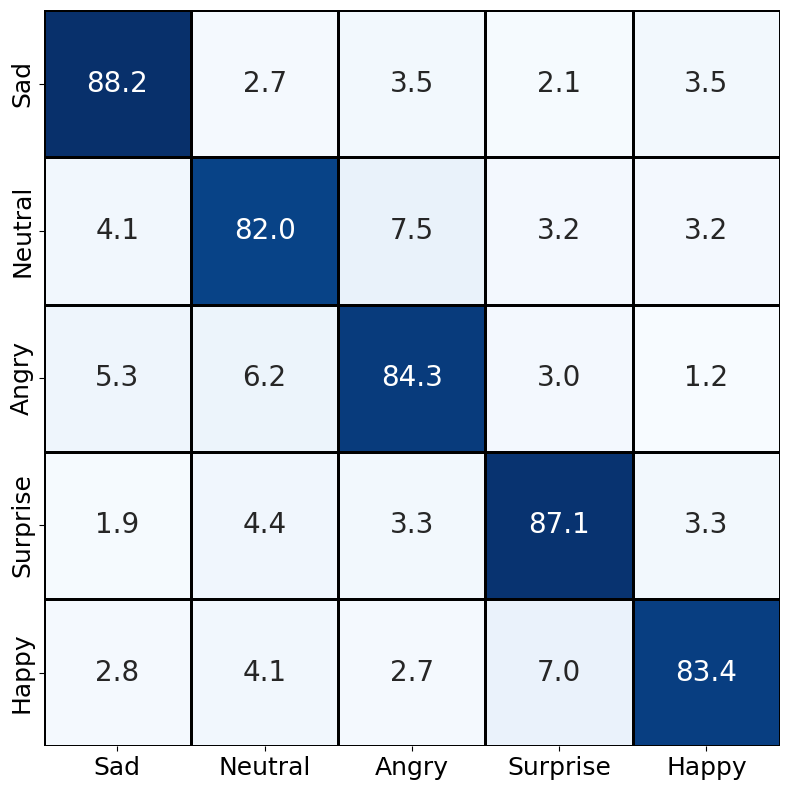}
        \caption{}
        \label{cmc}
    \end{subfigure}

    \vspace{0.3cm} % small vertical gap between rows

    \begin{subfigure}{0.3\columnwidth}
        \includegraphics[width=\linewidth]{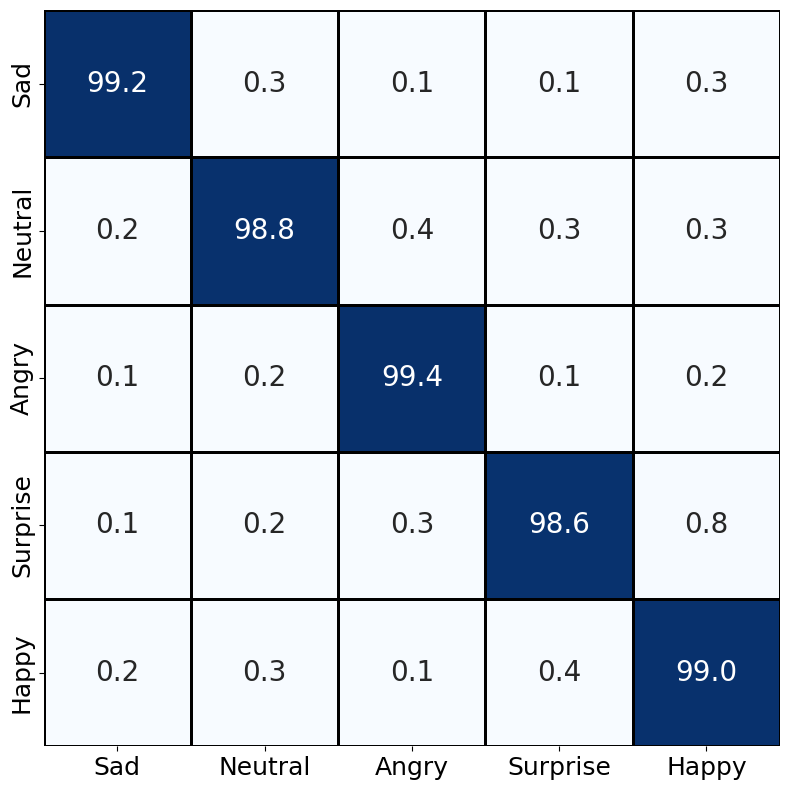}
        \caption{}
        \label{cmd}
    \end{subfigure}
    \hfill
    \begin{subfigure}{0.3\columnwidth}
        \includegraphics[width=\linewidth]{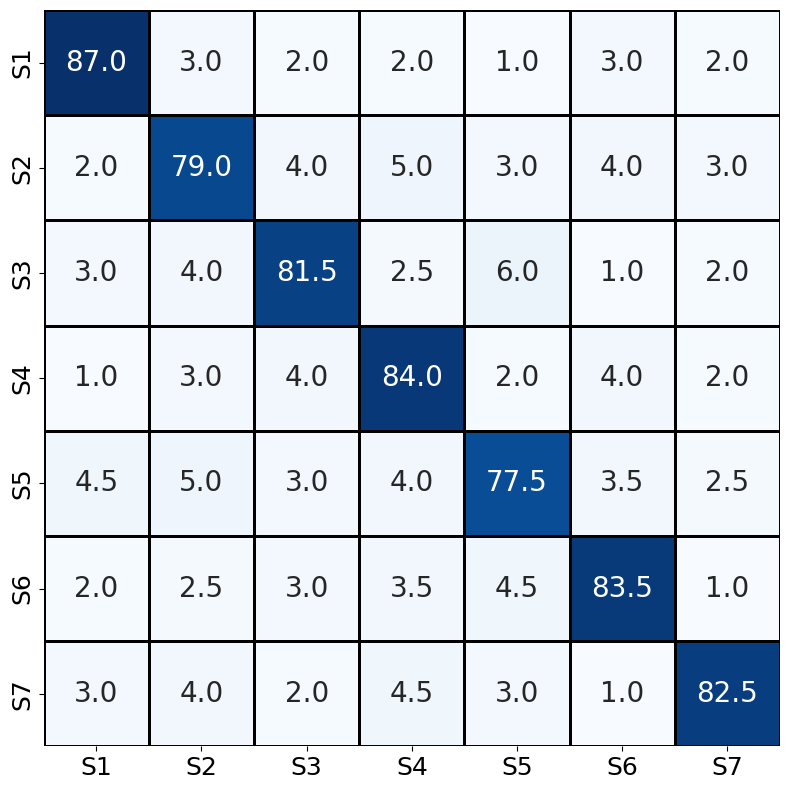}
        \caption{}
        \label{cme}
    \end{subfigure}
    \hfill
    \begin{subfigure}{0.3\columnwidth}
        \includegraphics[width=\linewidth]{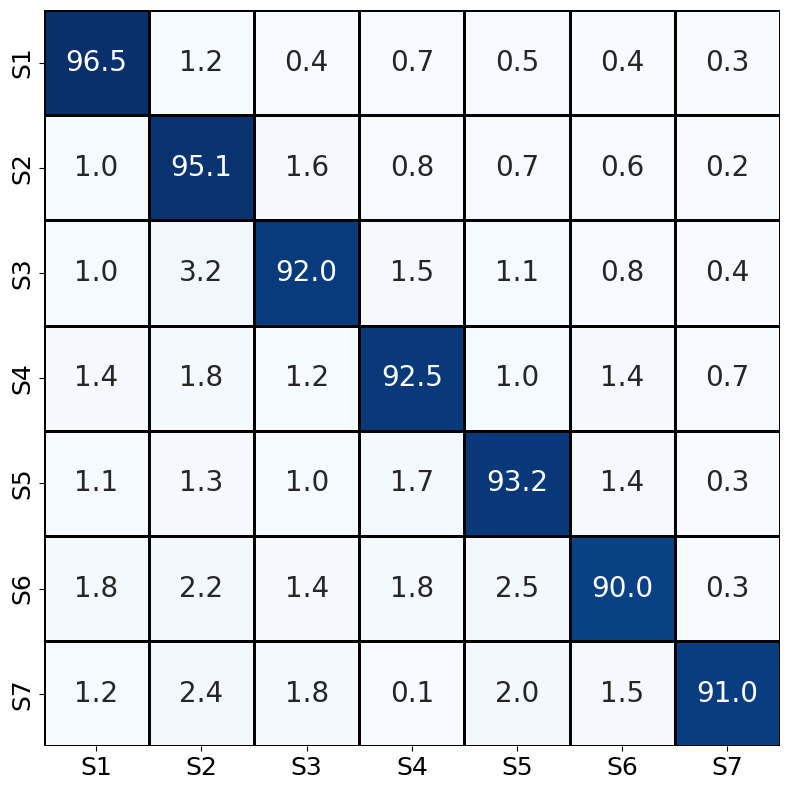}
        \caption{}
        \label{cmf}
    \end{subfigure}
    \vspace{0.5cm}
    \caption{
\textbf{Confusion matrices across emotion and manipulation targets on the English subset.} 
(\textbf{a}) OE classification using WavLM (singletask), 
(\textbf{b}) OE with \textbf{MiCuNet} using MMS + S-Mel (multitask), 
(\textbf{c}) CE with MMS + S-Mel (normal multitask), 
(\textbf{d}) CE with \textbf{MiCuNet} using MMS + S-Mel, 
(\textbf{e}) M classification using MMS + S-Mel (normal multitask), 
(\textbf{f}) M with \textbf{MiCuNet} using MMS + S-Mel. 
\textbf{MiCuNet} consistently improves class-wise separability, especially in multitask settings.
}
    \label{fig:cm}
\end{figure}
\begin{figure}[!t]
    \centering
    % First row: a–c (English subset)
    \begin{subfigure}[b]{0.3\columnwidth}
        \includegraphics[width=\linewidth]{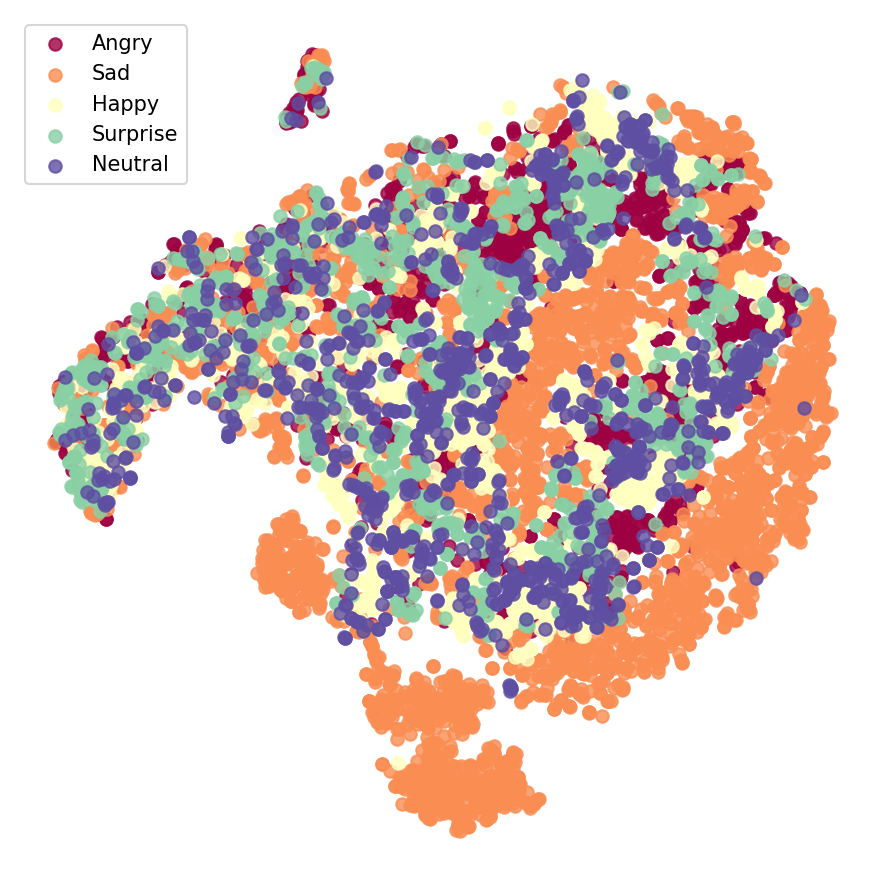}
        \caption{}
        \label{tsa}
    \end{subfigure}\hfill
    \begin{subfigure}[b]{0.3\columnwidth}
        \includegraphics[width=\linewidth]{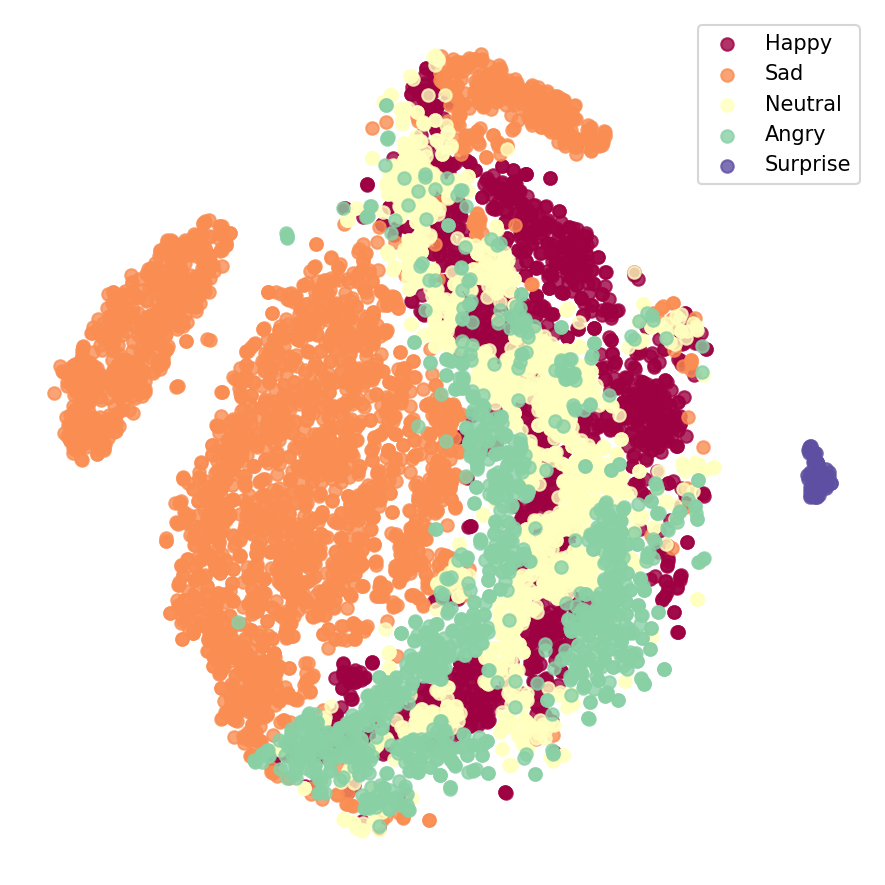}
        \caption{}
        \label{tsb}
    \end{subfigure}\hfill
    \begin{subfigure}[b]{0.3\columnwidth}
        \includegraphics[width=\linewidth]{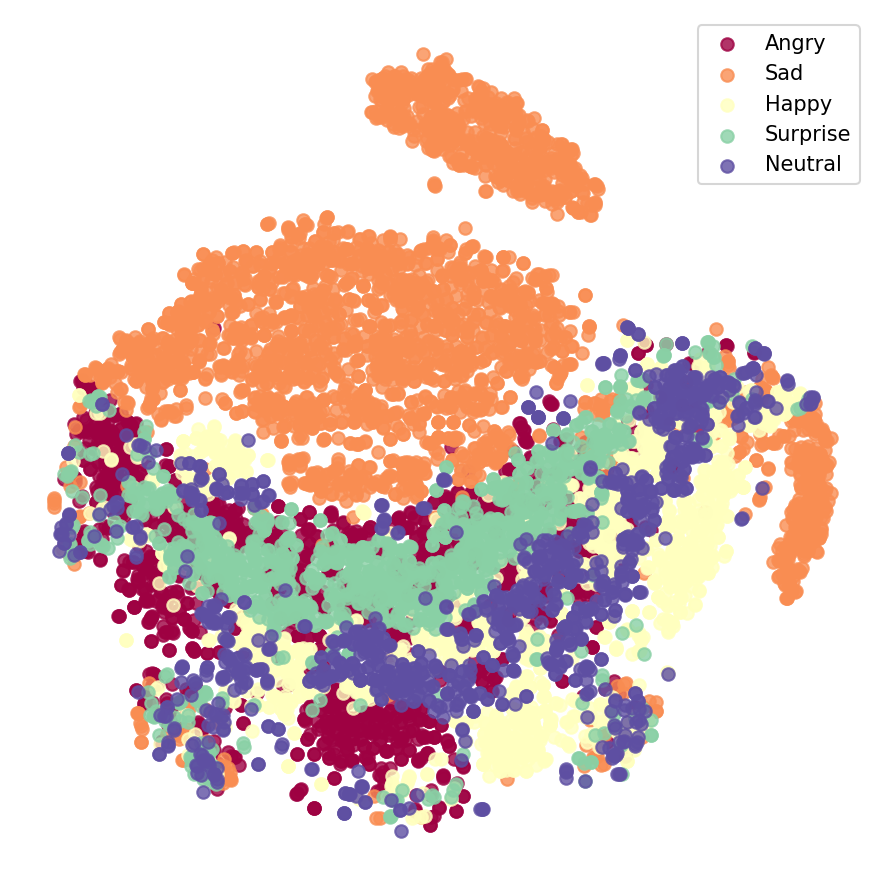}
        \caption{}
        \label{tsc}
    \end{subfigure}

      \vspace{0.3cm}

    % Second row: d–f (Chinese subset)
    \begin{subfigure}[b]{0.3\columnwidth}
        \includegraphics[width=\linewidth]{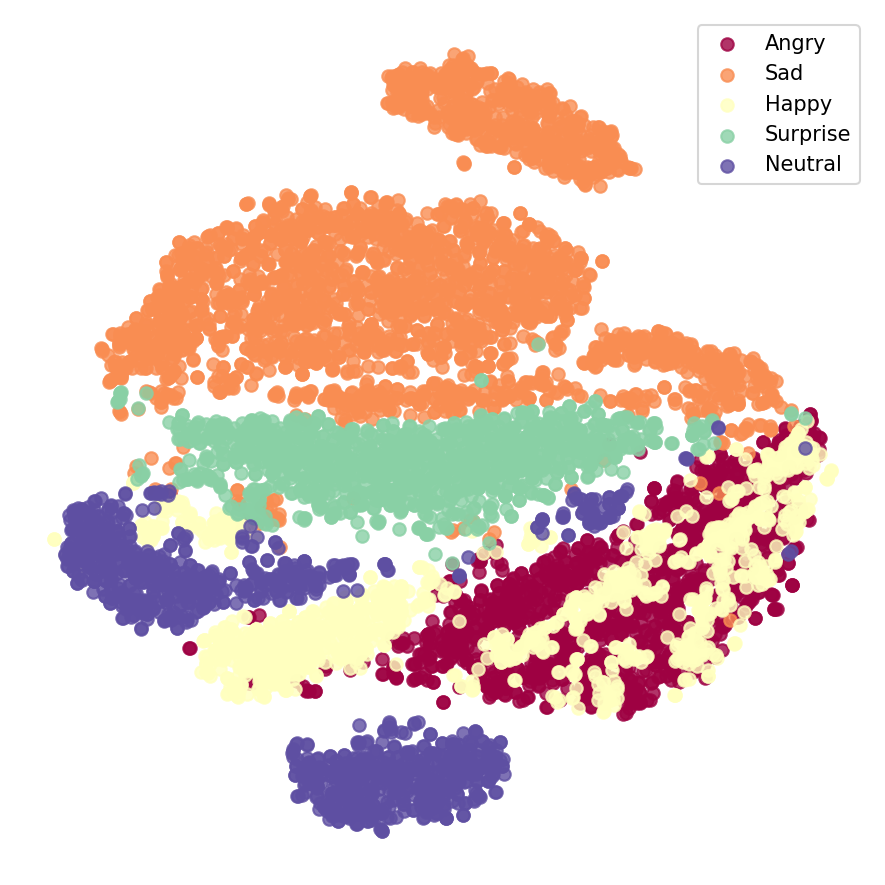}
        \caption{}
        \label{tsd}
    \end{subfigure}\hfill
    \begin{subfigure}[b]{0.3\columnwidth}
        \includegraphics[width=\linewidth]{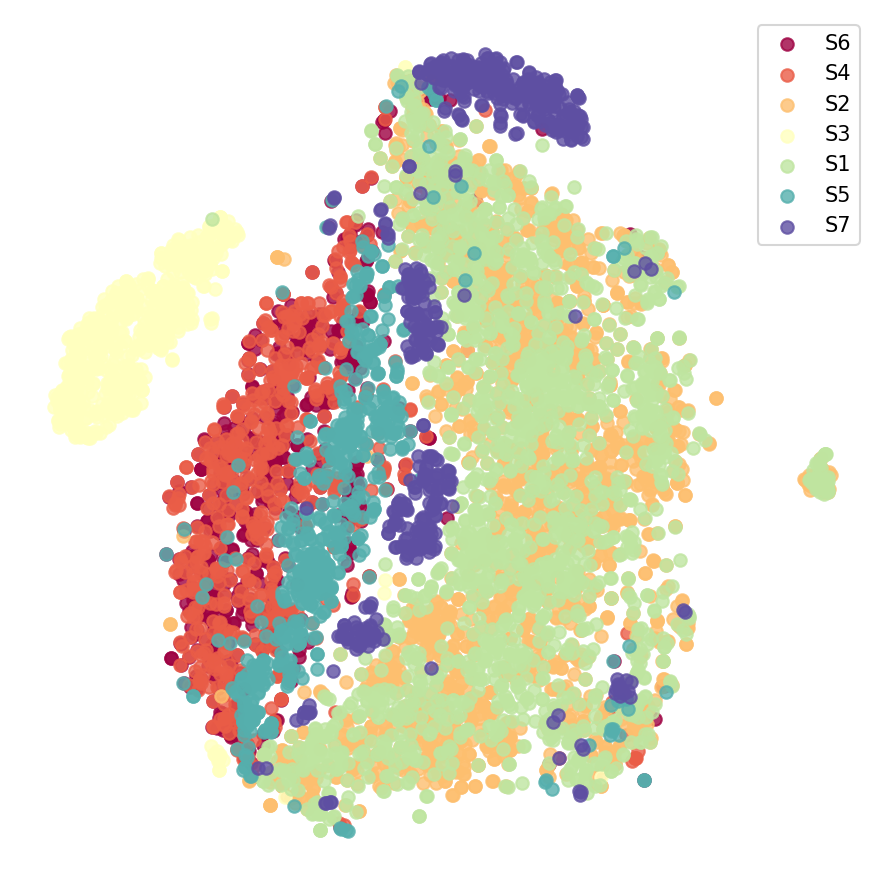}
        \caption{}
        \label{tse}
    \end{subfigure}\hfill
    \begin{subfigure}[b]{0.3\columnwidth}
        \includegraphics[width=\linewidth]{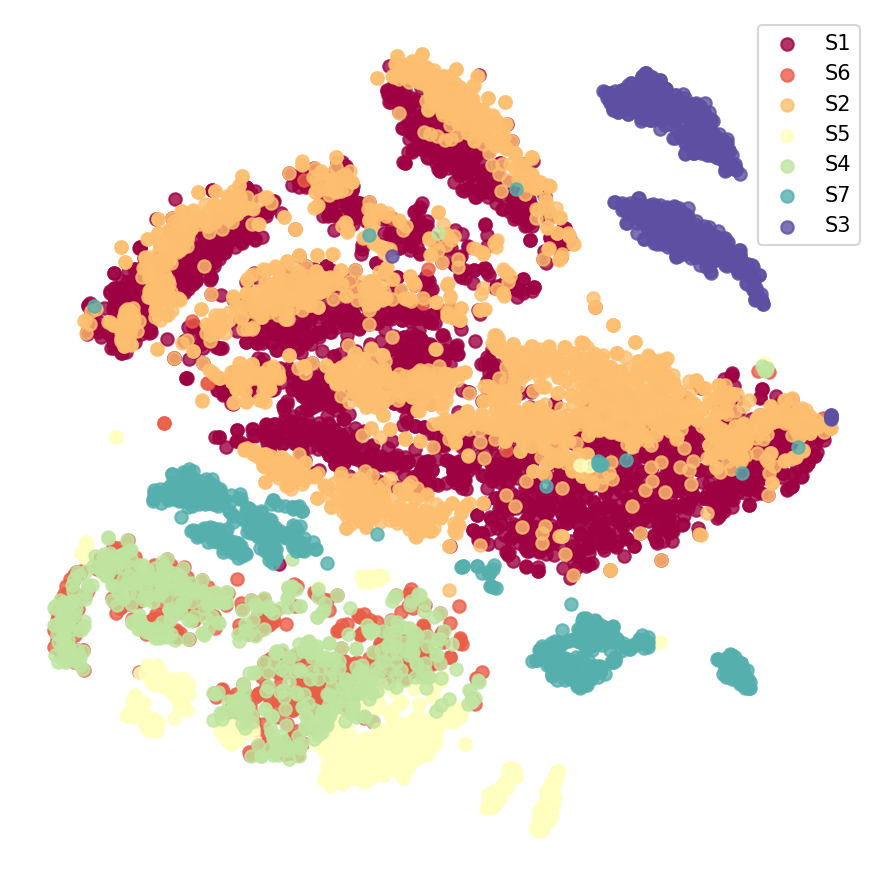}
        \caption{}
        \label{tsf}
    \end{subfigure}
  \vspace{0.5cm}

    % \caption{%
    %     t-SNE visualizations of learned embeddings. (a) x-vector (OE), (b) MMS (OE), (c) x-vector (CE), (d) MMS (CE), (e) x-vector (Model), (f) MMS (Model)
    % }
    \caption{t-SNE visualizations of learned embeddings. x-vector — (a) OE, (c) CE, (e) Model. \textbf{\texttt{MiCuNet}} using MMS + S-Mel — (b) OE, (d) CE, (f) Model.}
    \label{fig:tsne}
\end{figure}
\noindent \textbf{Visualization Analysis}: We present confusion matrices and t-SNE visualizations to qualitatively assess model behaviour across OE, CE, and M targets on the English subset. Figure~\ref{fig:cm} shows confusion matrices for both single-task and multi-task settings. Subfigures~\ref{cma}, \ref{cmc}, and \ref{cme} display baseline results using WavLM and MMS, while~\ref{cmb}, \ref{cmd}, and \ref{cmf} correspond to the proposed \textbf{\texttt{MiCuNet}} with MMS + S-Mel. To better understand how different representations behave across the OE, CE, and M tasks, we present t-SNE visualizations of the fused embedding space on the English subset in Figure~\ref{fig:tsne}. Subfigures~\ref{tsa}, \ref{tsc}, and \ref{tse} correspond to x-vector embeddings, while~\ref{tsb}, \ref{tsd}, and \ref{tsf} show results from the proposed \textbf{\texttt{MiCuNet}} using MMS + S-Mel fusion. Compared to x-vector, the \textbf{\texttt{MiCuNet}} representations exhibit clearer cluster boundaries and more compact class-wise distributions across all targets, suggesting that the geometry-aware fusion effectively enhances task-relevant separability in the learned feature space. Model refers to the manipulation source—the specific EVC generators.  \newline
\noindent \textbf{Evaluation Metrics:} We evaluate each task using Accuracy, macro-F1, and Equal Error Rate (EER). Since EER is initially defined for binary classifications, we extend it to multi-class settings using a macro one-vs-rest formulation. 

\subsection{Ablation Study}
To assess the contribution of individual geometric components and the learnable gating mechanism in \textbf{\texttt{MiCuNet}}, we conduct targeted ablation experiments. The results are summarized in Table~\ref{tab:4},\ref{tab:5},\ref{tab:6}.
\noindent \textbf{Impact of Geometric Manifolds:}
We evaluate the importance of each geometric space by removing the Hyperbolic (\texttt{–H}) or Spherical (\texttt{–S}) branch from the fusion module. Excluding either leads to performance drops across tasks and languages, with the Hyperbolic space having a stronger impact. When both are removed (\texttt{–H –S}), leaving only the Euclidean branch, performance degrades further—highlighting the value of curved manifolds in capturing emotional and manipulation patterns.
\noindent \textbf{Effect of Gating Mechanism:}
We further evaluate the role of the learnable gating module by replacing it with uniform averaging over the three manifolds (\texttt{No Gating}). Table~\ref{tab:4} summarizes the impact of each architectural component on performance.

\begin{table}[ht]
\setlength{\tabcolsep}{4.5pt}
\centering
\scriptsize
\begin{tabular}{l|ccc|ccc}
\toprule
\multirow{2}{*}{\textbf{Configuration}} 
  & \multicolumn{3}{c|}{\textbf{E → E}} 
  & \multicolumn{3}{c}{\textbf{C → C}} \\
\cmidrule(lr){2-4} \cmidrule(lr){5-7}
  & \textbf{OE} & \textbf{CE} & \textbf{M} 
  & \textbf{OE} & \textbf{CE} & \textbf{M} \\
\midrule
\textbf{\texttt{MiCuNet}}      & \textbf{0.66} & \textbf{0.31} & \textbf{0.51} & \textbf{0.78} & \textbf{0.40} & \textbf{0.70} \\
\midrule
Without Spherical (–S)           & 0.81 & 0.39 & 0.59 & 1.17 & 0.67 & 0.79 \\
Without Hyperbolic (–H)          & 0.88 & 0.46 & 0.63 & 1.28 & 0.72 & 0.86 \\
Only Euclidean (–H –S)           & 1.68 & 1.08 & 0.51 & 1.78 & 1.02 & 1.55 \\
No Learnable Gating              & 2.76 & 2.41 & 2.59 & 3.20 & 3.63 & 3.80 \\
\bottomrule
\end{tabular}
\caption{Ablation study on \texttt{MiCuNet} using MMS + S-MEL showing Equal Error Rate (\%) for Original Emotion (OE), Current Emotion (CE), and Manipulation Source (M).}
\label{tab:4}
\end{table}

\noindent \textbf{Generalization to Unseen EVC Sources:}
To evaluate the robustness of \textbf{\texttt{MiCuNet}} to previously unseen manipulation methods, we conduct a leave-two-out evaluation across the seven emotional voice conversion (EVC) models. In each case, the model is trained on speech generated by five EVC converters and tested exclusively on samples from the remaining two (S6 and S7), which are held out during training. Table~\ref{tab:5} reports performance for both Original Emotion (OE) and Current Emotion (CE) predictions within the same language—English (E→E) and Chinese (C→C). Despite never encountering S6 or S7 during training. Table~\ref{tab:6} extends this evaluation to cross-lingual settings, where models trained on one language are evaluated on another (E→C and C→E).

\begin{table}[ht]
\setlength{\tabcolsep}{4pt}
\centering
\scriptsize
\begin{tabular}{l|cc|cc|cc|cc}
  \toprule
  & \multicolumn{4}{c|}{\textbf{E → E}} & \multicolumn{4}{c}{\textbf{C → C}} \\
  \cmidrule(lr){2-5} \cmidrule(lr){6-9}
  \textbf{Model} 
    & \multicolumn{2}{c|}{\(\mathbf{S6}\)} 
    & \multicolumn{2}{c|}{\(\mathbf{S7}\)} 
    & \multicolumn{2}{c|}{\(\mathbf{S6}\)} 
    & \multicolumn{2}{c}{\(\mathbf{S7}\)} \\
  \cmidrule(lr){2-3} \cmidrule(lr){4-5} \cmidrule(lr){6-7} \cmidrule(lr){8-9}
  & OE & CE & OE & CE & OE & CE & OE & CE \\
  \midrule
  MMS + S–MEL 
    & 1.78 & 1.20 & 1.69 & 1.21 
    & 1.84 & 1.39 & 1.77 & 1.26 \\
  \bottomrule
\end{tabular}
\caption{EER (\%) of \texttt{MiCuNet} on unseen EVC models (S6, S7) for OE and CE within the same language.}
\label{tab:5}
\end{table}

\begin{table}[!h]
\setlength{\tabcolsep}{4pt}
\scriptsize
\centering
\begin{tabular}{l|cc|cc|cc|cc}
\toprule
               & \multicolumn{4}{c|}{\textbf{E → C}}                                        & \multicolumn{4}{c}{\textbf{C → E}}                                        \\
\cmidrule(lr){2-5} \cmidrule(lr){6-9}
\textbf{Model} & \multicolumn{2}{c|}{\(\mathbf{S6}\)} & \multicolumn{2}{c|}{\(\mathbf{S7}\)} & \multicolumn{2}{c|}{\(\mathbf{S6}\)} & \multicolumn{2}{c}{\(\mathbf{S7}\)} \\
\cmidrule(lr){2-3} \cmidrule(lr){4-5} \cmidrule(lr){6-7} \cmidrule(lr){8-9}
               & OE               & CE               & OE                & CE               & OE                & CE              & OE               & CE               \\
\midrule
MMS + S–ME     & 1.95             & 1.88             & 1.91              & 1.56             & 2.31              & 2.00            & 2.17             & 1.98             \\
\bottomrule
\end{tabular}
\caption{Cross-lingual EER (\%) of \texttt{MiCuNet} on unseen EVC models (S6, S7) for OE and CE, with training and testing across different languages.}
\label{tab:6}
\end{table}
% \vspace{-2.5mm}
\section{Conclusion}
In this work, we addressed the task of tracing of emotional and manipulation attributes in synthetically generated speech. We introduced \texttt{\textbf{MiCuNet}}, a multitask framework that fuses speech foundation model (SFM) embeddings with spectrogram-based auditory features using a mixed-curvature projection strategy. By mapping representations across hyperbolic, Euclidean, and spherical spaces with a learnable gating mechanism, \texttt{MiCuNet} effectively captures complementary cues critical for emotion and manipulation source inference. Evaluations on the EmoFake dataset across English and Chinese demonstrate that \texttt{\textbf{MiCuNet}} consistently outperforms conventional fusion baselines on all tasks, including original emotion, current emotion, and manipulation source prediction. Our findings underscore the value of using curvature-aware representations for understanding synthetic speech. \newline
% \vspace{-0.8mm}
\section*{Ethical Statement}
This work is conducted with the intent to improve transparency, accountability, and forensic analysis in the context of synthetic speech technologies. While the techniques developed in this study can detect and trace emotional manipulation in speech, we acknowledge that such technologies may also raise ethical concerns around surveillance, privacy, and potential misuse. All experiments were performed on publicly available datasets, and no human subjects were directly involved.
\section*{Limitation}
The most significant limitation of this study lies in the lack of publicly available datasets explicitly designed for emotional manipulation in synthetic speech. To the best of our knowledge, EmoFake is the only publicly available dataset that includes parallel annotations for original emotion, manipulated emotion, and the conversion source. This scarcity of data limits the ability to perform broader benchmarking and evaluate cross-dataset generalization in more diverse and realistic settings. This makes comprehensive benchmarking and cross-dataset generalization challenging at this stage.

\bibliography{main}
\end{document}